%% file: Main.tex
\documentclass[journal,transmag]{IEEEtran}
\ifCLASSINFOpdf
\else
\fi
 \usepackage{booktabs}
 \usepackage{multirow}
\usepackage{slashbox}
\usepackage{diagbox}
\usepackage{graphicx}
\usepackage{subcaption}
\usepackage{xcolor}
\usepackage{authblk}
\usepackage{subfiles}
\usepackage{amsmath, amsthm, amssymb}
\usepackage{tikz,pgfplots}
\usepackage[inline]{enumitem}

\usetikzlibrary{positioning}
\pgfplotsset{
    compat=newest,
    /pgfplots/legend image code/.code={%
        \draw[mark repeat=2,mark phase=2,#1] 
            plot coordinates {
                (0cm,0cm) 
                (0.1cm,0cm)
                (0.1cm,0cm)
                (0.1cm,0cm)
                (0.4cm,0cm)%
            };
    },
}

\begin{document}
%
\title{Generalized Tensor Summation Compressive Sensing Network (GTSNET): An Easy to Learn Compressive Sensing Operation}


\author[1]{Mehmet Yama\c{c}}
\author[1]{Ugur Akpinar}
\author[1]{Erdem Sahin}
\author[2]{Serkan Kiranyaz}
\author[1]{Moncef Gabbouj}
\affil[1]{Tampere University, Faculty of Information Technology and Communication Sciences, Tampere, Finland}
\affil[2]{Department of Electrical Engineering, Qatar University, Qatar}%

\markboth{Journal of \LaTeX\ Class Files,~Vol.~14, No.~8, August~2021}%
{Shell \MakeLowercase{\textit{et al.}}: Bare Demo of IEEEtran.cls for IEEE Transactions on Magnetics Journals}
%



\maketitle

\begin{abstract}
In the compressive sensing (CS) literature, the efforts can be divided into two groups: finding a measurement matrix that preserves the compressed information at the maximum level, and finding a robust reconstruction algorithm for the compressed information. In the traditional CS setup, the measurement matrices are selected as random matrices, and optimization-based iterative solutions are used to recover the signals. However, when we handle large or multi-dimensional signals, using such random matrices become cumbersome especially when it comes to iterative optimization-based solutions. Even though recent deep learning-based solutions boost the reconstruction accuracy performance while speeding up the recovery, still jointly learning the whole measurement matrix is a difficult process. This is why recent state-of-the-art deep learning CS solutions such as convolutional compressive sensing network (CSNET) use block-wise CS schemes to ease the learning of CS operation. In this work, we introduce a separable multi-linear learning of the CS matrix by representing it as the summation of arbitrary number of tensors. For a special case where the CS operation is set as a single tensor multiplication, the model is reduced to the learning-based separable CS; while a dense CS matrix can be approximated and learned as the summation of multiple tensors. Both cases can be used in CS of two or multi-dimensional signals e.g., images, multi-spectral images, videos, etc. Structural CS matrices can also be easily approximated and learned in our multi-linear separable learning setup with structural tensor sum representation. Hence, our learnable generalized tensor summation compressive sensing operation encapsulates most CS setups including separable CS, non-separable CS (traditional vector-matrix multiplication), structural CS, and CS of the multi-dimensional signals. Even though the proposed method is suitable for one, two, or multi-dimensional signals, for the case study we worked on gray-scale and RGB images. Compared to the block-wise CS scheme, tensorial learning eases the problem of blocking artifacts, and leads to a superior performance. For both gray-scale and RGB images, the proposed scheme surpasses most state-of-the-art solutions, especially in lower measurement rates. Although the performance gain remains limited from tensor to the sum of tensor representation for gray-scale images, it becomes significant in the RGB case. The software implementation of the proposed network is publicly shared at https://github.com/mehmetyamac/GTSNET.  

\end{abstract}



%
\IEEEpeerreviewmaketitle

\section{Introduction}

\IEEEPARstart{C}{ompressive} sensing (CS) theory has attracted a lot of attention since its first appearance in 2005 \cite{CS1}. CS theory claims that a signal can be sampled with far fewer measurements than the conventional Nyquist/Shannon-based sampling methods use. It has been applied in many fields such as CS-based MRI imaging \cite{MR}, radar monitoring systems \cite{radar1, radar2}, and ECG measurements in a health monitoring system \cite{ecg}. Along with sampling, the technology has been adopted in many other fields. For instance,  in a conventional CS system, random or pseudo-random measurement matrices are used, enabling a CS-based encryption mechanism \cite{CSinEnc, yamacEn}.

In Nyquist/Shannon based data acquisition systems, the reconstruction process is performed by sinc interpolation, which is a linear process and does not require expensive computations. The traditional CS-based data acquisition systems require advanced optimization-based iterative algorithms such as $\ell_1$-minimization techniques \cite{BP, BPDN, dantzig2}. Even if convex relaxation can bring a guarantee of sparse recovery with polynomial time, most solvers work in an iterative manner, and it makes them infeasible for real-time applications especially for large-scale signals, such as vectorized images. Moreover, $\ell_1$ type estimators may lead to an unbiased estimation of the sparse signal \cite{unbias}. There have been significant efforts spent to have faster recovery algorithms such as \cite{TVAL3, DAMP, figueiredo2007gradient} which are more feasible for a CS imaging system or similar multi-dimensional signals. However, the optimization-based recovery in a sparse domain can completely fail under some measurement rates, which are determined by the phase transition of the algorithms \cite{phasetransition}. Moreover, the signal of interest in real applications rarely becomes strictly sparse in any sparsifying domain.

The first category of the deep learning-based CS approaches includes the works that use neural networks only for the reconstruction part \cite{SDA,reconnet,ISTA-Net}. They generally use conventional random matrices as the CS operators. To handle the images (2D signal), they apply the CS matrices to the vectorized smaller blocks of the image of interest. The well-known state-of-the art examples of this category of work can be listed as stacked denoising autoencoder (SDA) \cite{SDA}, non-iterative reconstruction of the compressively sensed images using CNN (ReconNet) \cite{reconnet}, and learned version of iterative shrinkage thresholding algorithm for CS imaging (ISTA-Net) \cite{ISTA-Net}. Among them, SDA uses the fully connected layers while the others adopt convolutional layers in their network. As a reconstruction part, ReconNet introduces fully convolutional layers, and this is why it is a non-iterative recovery framework that significantly reduces the computational time. ISTA-Net is based on iterative soft thresholding algorithms, and can be put into the category of deep unrolling techniques.

The second category of deep learning attempts can be enlisted as the ones that jointly learn CS matrices and reconstruction part instead of using conventional CS matrices. The recent state-of-the-art networks in this category are convolutional compressive sensing network (CSNET) \cite{CSNET} and scalable convolutional compressive sensing network (SCSNET) \cite{SCSNET}. These works also handle CS of both gray-scale and RGB images in a block-by-block manner. However, they learn these CS matrices using convolutional kernels having the same size as the image blocks. In the reconstruction part they use convolutional layers to recover the full image as a whole. In this way, they can significantly improve the blocking artifacts. 

In this study, we propose a novel network, the so-called Generalized Tensor Summation Networks (GTSNETs), that can \textit{jointly} learn both CS matrix and reconstruction algorithms. Contrary to previous attempts, a GTSNET can generalize most of the sensing systems such as unconstraint CS matrices, separable CS matrices, and structural CS matrices. To learn unconstraint CS matrices (the matrices that can be applied over the whole vectorized image or any multidimensional signal), the matrices are factorized as the summation of $T$ tensor summation, which makes the learning of CS matrices for multi-dimensional signals such as images or videos feasible.  For the special case of $T=1$, the CS system reduces down to a separable CS system \cite{separableCS}, which is also known as Kronecker CS \cite{kronecker}.

In a GTSNET, CS operation can be performed directly over the spatial domain or in any other separable transformation basis like CS in the frequency domain using DCT. This is why GTSNET can generalize many CS systems, and thus we use the term Generalized Tensor Summation (GTS). When it comes to the performance comparison with a traditional deep learning approach, especially for lower measurement rates, the proposed system exhibits a superior performance in terms of PSNR and SSIM with a particular improvement over the fine details. At the same time, GTSNET performs signal reconstruction from compressively sensed measurements in a feed-forward manner and this significantly reduces the computational complexity compared to the iterative approaches.

The rest of the paper is organized as follows. In Section~\ref{Preliminaries}, we shall make a brief introduction to compressive sensing, separable and multidimensional. Then, the proposed learnable compressive sensing operations will be presented in Section~\ref{proposed}. In Section~\ref{results} extensive experimental results will be presented for the CS in both gray-scale and RGB images. We shall then present comparative evaluations in spatial and frequency domains. In addition, we shall investigate which information is more preserved when $T$ is increased. Finally, the conclusions are drawn in Section~\ref{conclusion}.


%
%
%
%


\section{Preliminaries and Prior Art}
\label{Preliminaries}
\subsection{Compressive Sensing}
CS \cite{CS1, CS2} theory has shown that a sparse signal can be recovered from far fewer measurements than traditional Shannon-Nyquist-based data acquisition methods use. Mathematically speaking, let a CS scheme linearly extracts $m$ number of measurements of the signal, $\mathbf{s} \in \mathbb{R}^N$, i.e.,
    \begin{equation}
 \mathbf{y}= \mathbf{\Psi s} \label{CS},
\end{equation}
where the measurement matrix, $\mathbf{\Psi} \in \mathbb{R}^{m \times N}$ represents the linear data acquisition with $m << N$.
In the CS literature, the efforts of designing such a linear measurement system can be categorized into two groups: (i) Finding a measurement matrix, $\mathbf{\Psi}$ which maximally preserves the information of $\mathbf{s}$ while transforming it in a lower-dimensional subspace as in Eq.~\eqref{CS}. (ii) Finding a robust reconstruction algorithm, which is able to recover $\mathbf{s}$ from $\mathbf{y}$ in a reasonable time with a tolerable reconstruction error. 

From elementary linear algebra, one can easily say that Eq.~\eqref{CS} is an underdetermined linear system of equations where for a given $\mathbf{\Psi}$ and $\mathbf{y}$ pair, $\mathbf{s}$ has infinitely many solutions. Therefore, at least one more assumption is needed to have unique solution. For instance, if we know that the signal of interest, $\mathbf{s}$, is sparse in a proper sparsifying domain $\Phi$, then Eq.~\eqref{CS} can be expressed as
\begin{equation}
     \mathbf{y} = \mathbf{\Psi} \mathbf{\Phi} \mathbf{x} = \mathbf{A x} \label{CS2},
\end{equation}
where $\mathbf{x} \in \mathbb{R}^N$ is sparse or compressible coefficient vector (e.g., if it is $k$-sparse $\left \| \mathbf{x} \right \|_{\ell_0^N} \leq k$) and $\mathbf{A} = \mathbf{\Psi \Phi}$, which can be named as equivalent dictionary \cite{equivalent}. Under the assumption that the coefficient vector is $k$-sparse, then the following sparse representation, 
\begin{equation}
\min_{\mathbf{x}} ~ \left \| \mathbf{x} \right \|_{0}~ \text{subject to}~ \mathbf{Ax}=\mathbf{y} \label{sparse_rep}
\end{equation}
is unique if $m \geq 2k$ and the minimum number of linearly dependent columns of $\mathbf{A}$ (see the definition of spark of a matrix \cite{spark}) is greater than $2k$ \cite{spark}. However, the problem in Eq.~\eqref{sparse_rep} is non-convex and known to be NP-hard. Fortunately, the most common approach will be the relaxation of it to an $\ell_1$ minimization problem,
\begin{equation}
     \arg \min_{\mathbf{x}} \left \| \mathbf{x} \right \|_1 ~s.t. ~ \mathbf{x} \in \mho \left ( \mathbf{y} \right ) \label{Eq:l1}
\end{equation}
where $\mho \left ( \mathbf{y} \right ) = \left \{ \mathbf{x}: \mathbf{Ax}=\mathbf{y} \right \}$ in noisy-free case, which is known as Basis Pursuit (BP) \cite{BP}. To guarantee the equivalence of the solutions of Eq.~\eqref{sparse_rep} and Eq.~\eqref{Eq:l1}, some properties of $\mathbf{A}$ are needed such as Null Space Property (NSP) \cite{NSP1, NSP2}. NSP can also be used to deal with approximately sparse signals. Moreover, if one should deal with approximately sparse signals in a noisy environment, a stronger property known as Restricted Isometry Property (RIP) \cite{RIP1, candesRIP} can be borrowed from the CS literature. In this noisy case, the constraint in the optimization problem can be relaxed by setting $\mho \left ( \mathbf{y} \right ) = \left \{ \mathbf{x}: \left \| \mathbf{ Ax}-\mathbf{y}  \right \|_2  \leq \epsilon \right \}$ which is known as Basis Pursuit Denoising (BPDN) \cite{BPDN} or Dantzig Selector \cite{dantzig} if we set  $\mho \left ( \mathbf{y} \right ) = \left \{ \mathbf{x}: \left \| \mathbf{A}'\left ( \mathbf{y}-\mathbf{Ax} \right ) \right \|_{\infty } \leq \lambda   \right \}$. Although RIP can be used for both stability and uniqueness analysis, the calculation of Restricted Isometric Constant (RIC) of $\mathbf{A}$ (defined with RIP of $\mathbf{A}$) generally requires a combinatorial search. Therefore, instead of RIC of a matrix, another important measure of a measurement matrix is defined in the literature. This is a functional $\mu (\mathbf{A}) = \max_{i,j} \left | A_{i,j} \right |$, which is known as coherence. In CS literature, choosing the best measurement matrix $\mathbf{\Psi}$ according to sparsifying matrix $\mathbf{\Phi}$ is well studied in terms of $\mu(\mathbf{A})$. The system in Eq.~\eqref{CS2} is nothing but a linear dimensional reduction system. To be able to preserve enough information while transforming $\mathbf{x}$ to $\mathbf{y}$, we generally wish each row of matrix $\mathbf{A}$ to get enough information from each element of $\mathbf{x}$. In other words, the flatness of the rows of $\mathbf{A}$ is desired. This can be satisfied when the rows of the measurement matrix $\mathbf{\Psi}$ is not sparse in $\mathbf{\Phi}$. To  describe  this  concept,  the  functional called ”mutual coherence” is defined, 
\begin{equation}
    \mu \left ( \mathbf{\Psi}, \mathbf{\Phi} \right ) := \max_{1 \leq k \leq m, 1 \leq j \leq N } \left | \left \langle \psi_k, \phi_j \right \rangle \right |
\end{equation}
which measures the coherence between $\mathbf{\Psi}$ and $\mathbf{\Phi}$, where $\psi_k$ is the $k$'th row of $\mathbf{\Psi}$ and $\phi_j$ is the $j$'th column of $\mathbf{\Phi}$. It is clear that $\mu (\mathbf{\Psi}, \mathbf{\Phi} ) \in \left [ \frac{1}{\sqrt N}, 1 \right ]$ when $\Psi$ has normalized rows and $\Phi$ has normalized columns. A theoretical lower bound to guarantee exact recovery for BP on the number of measurement in terms of defined mutual coherence can be found in \cite{Incoherence,Sapiro} as,
 \begin{equation}
m \geq \kappa .N. \mu^2 (\mathbf{\Psi}, \mathbf{\Phi} ).k .\log N,
 \end{equation}
 where $\kappa$ is a positive constant.
In plain terms, the minimum number of required measurements is dictated by the mutual coherence, and one wishes to keep it a minimum in a CS system. For instance, the measurement matrices, which have random waveforms with \textit{i.i.d} elements such as Gaussian, are well known to be incoherent with any fixed basis, i.e., $\mu\left ( \mathbf{\Psi}, \mathbf{\Phi} \right )     \approx  \frac{\sqrt{2 \log (N)}}{\sqrt{N}}$ \cite{MC_random}.    

\subsection{Multi-dimensional and Separable Compressive Sensing}
As reviewed above, the mathematical foundation of the conventional CS scheme is well established. However, this traditional scheme, where dimensional reduction is performed via vector-matrix multiplication and recovery is represented via $\ell_1$-minimization, may not be convenient in most multi-dimensional signal acquisition schemes such as compressive sensing of imaging systems. For instance, assume that the signal of interest is a $512 \times 512$ gray-scale image, $\mathbf{S}$. Assume that we wish to build a CS system with a measurement rate of $\frac{m}{N} = 0.36$. In that CS scheme that samples the vectorized image, $\mathbf{s} \in \mathbb{R}^N$ with $N=512^2$, the measurement matrix size will be $m \times N = 94372 \times 262 144$. The conventional CS recovery algorithms such as $\ell_1$-minimization techniques are iterative algorithms and in each iteration, they perform matrix-vector multiplications using  CS matrix and the transpose of it. However, even saving alone such massive size matrices requires more than 80GB of storage. Therefore, the computational complexity of the iterative recovery algorithms becomes cumbersome. As a remedy, block-base CS and separable CS imaging \cite{separableCS} have become the most frequently used approaches. Among them, separable CS (also known as Kronecker CS) has the advantage of introducing fewer  blocking artifacts. In a separable CS imaging introduced in \cite{separableCS}, the CS sampling operator is separable over horizontal and vertical axes, i.e., $ \mathbf{Y} = \mathbf{\Psi_1} \mathbf{S} \mathbf{\Psi_2'}$, where $\mathbf{S} \in \mathbb{R}^{\sqrt{N} \times \sqrt{N} }$ is the input image, $\mathbf{s} \in \mathbb{R}^N$, in its original matrix form, and $\mathbf{\Psi_1} \in \mathbb{R}^{\sqrt{m} \times \sqrt{N} }$  and $ \mathbf{\Psi_2} \in \mathbb{R}^{\sqrt{m} \times \sqrt{N} } $ are the measurement matrices. In that way, the computational cost of the matrix multiplications is reduced from $2 \times m \times N$ flops to $4 \times \sqrt{m} \times \sqrt{N}$ flops compared to conventional CS setup. Moreover, this separable CS setup can be easily formulated in a traditional CS setup, which makes the analysis and algorithms of CS theory still valid. For instance, consider that the sparsifying basis is also separable as in 2D DCT matrices, then CS in matrix-vector form is nothing but $vec(\mathbf{Y}) = \mathbf{\Psi_1} \otimes \mathbf{\Psi_2}~ vec(\mathbf{S})  = \mathbf{A_1} \otimes \mathbf{A_2}~ vec(\mathbf{X})$, where $\mathbf{A_i} = \mathbf{\Psi_i} \mathbf{\Phi_i}$, $\mathbf{X} \in \mathbb{R}^{\sqrt{N} \times \sqrt{N} }$ is a sparse coefficient matrix and $\otimes$ is the Kronecker product. Let us assume the separable measurement matrices are Gaussian projection matrices, then mutual coherence between $\mathbf{\Psi_1} \otimes \mathbf{\Psi_2}$ and $\mathbf{\Phi_1} \otimes \mathbf{\Phi_2}$ can be easily calculated, i.e., $\mu \left ( \mathbf{\Phi_1} \otimes \mathbf{\Phi_2}, \mathbf{\Psi_1} \otimes \mathbf{\Psi_2} \right ) \approx \frac{log (N)}{\sqrt{N}}$. Hence, the mutual coherence increases $\sqrt{\frac{1}{2} log(N)}$ times and the number of the necessary measurement increases by the square of it compared to a conventional setup where CS matrix is unconstrained Gaussian projection matrix. \textit{That is to say, in a separable CS setup, although computational complexity decreases, the minimum number of required measurements increases as a trade-off compared to conventional unconstrained CS scheme.  }

In general multi dimensional and separable CS setup, the $J$-dimensional signal, $\mathbf{\mathcal{S}} \in \mathbb{R}^{n_1 \times n_2 ...\times n_J}$ with $N = \prod_{j=1}^J n_j$, can be acquired by separable sensing operator: 
\begin{equation}
    \mathcal{Y} = \mathbf{\mathcal{S}} \times_1 \mathbf{\Psi_1} \times_2 \mathbf{\Psi_2} ... \mathbf{\Psi_{J-1}} \times_J \mathbf{\Psi_J}, \label{MDCS}
\end{equation}
where $\mathbf{\mathcal{S}} \times_i \mathbf{\Psi_i}$ is the i-mode product of tensor $\mathbf{\mathcal{S}}$ and matrix $\mathbf{\Psi_i} \in \mathbb{R}^{m_i \times n_i}$ and $\mathbf{\mathcal{Y}} \in \mathbb{R}^{m_1 \times m_2 ...\times m_J}$ is CS tensor, with $m = \prod_{j=1}^J m_j$. Assuming that the sparsifying basis is also separable, then Eq.~\eqref{MDCS} can be re-cast as,
\begin{equation}
    \mathcal{Y} = \mathbf{\mathcal{X}} \times_1 \mathbf{A_1} \times_2 \mathbf{A_2} ... \mathbf{A_J} \times_J \mathbf{A_J} \label{MDCS2}
\end{equation}
where $\mathbf{A_i} = \mathbf{\Psi_i} \mathbf{\Phi_i}$ and $\mathbf{\mathcal{X}} \in \mathbb{R}^{n_1 \times n_2 ...\times n_J}$ is the sparse representation tensor. Eq.~\eqref{MDCS2} can be cast as a vector-matrix multiplication, 
\begin{equation}
    \mathbf{y} = \left ( \mathbf{A_1} \otimes \mathbf{A_2} \otimes ...\otimes \mathbf{A_J} \right ) \mathbf{x},
\end{equation}
where $\mathbf{y} = vec(\mathcal{Y})$ and $\mathbf{x} = vec(\mathcal{X})$. Therefore, the conventional CS recovery techniques defined in Eq.~\eqref{Eq:l1} can be used and this setup is also known as tensor compressive sensing or Kronecker compressive sensing \cite{kronecker}.  


\section{Generalized Structural Tensor Sum Compressive Sensing}
\label{proposed}
\subsection{Tensor sum as a computationally efficient approximation of non-separable CS Matrix }
\label{sec:TensorSumCoherence}
\begin{figure}[h]
 \centering
  \includegraphics[width=0.245\textwidth]{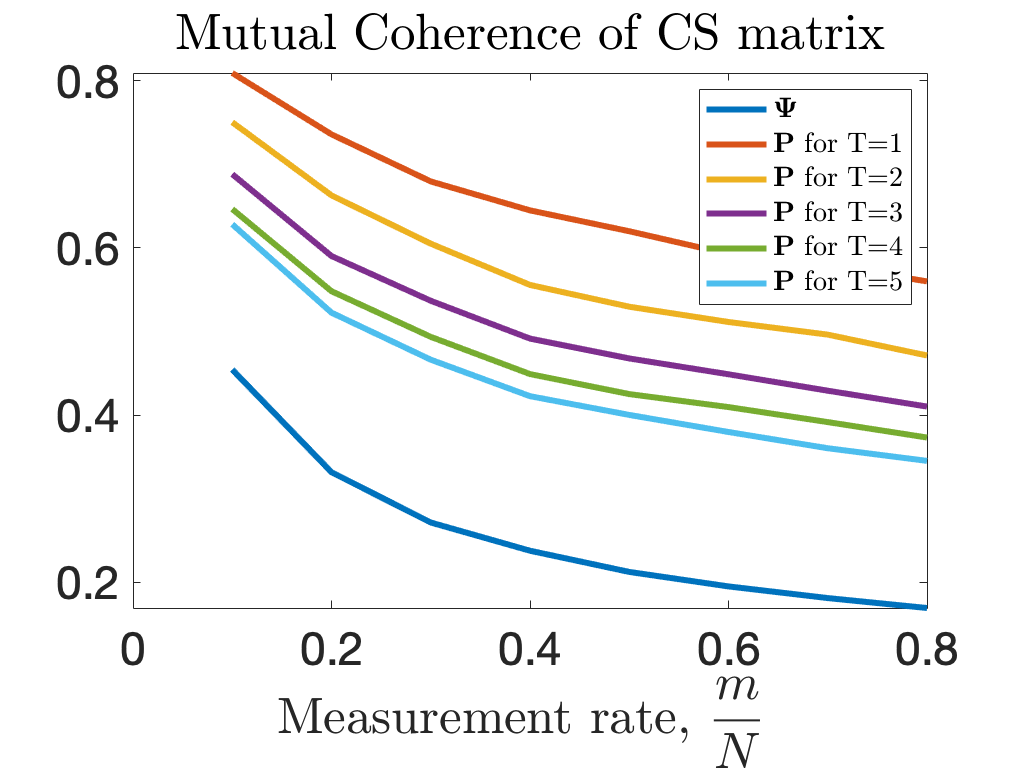}
\hspace{-0.25cm}
 \includegraphics[width=0.245\textwidth]{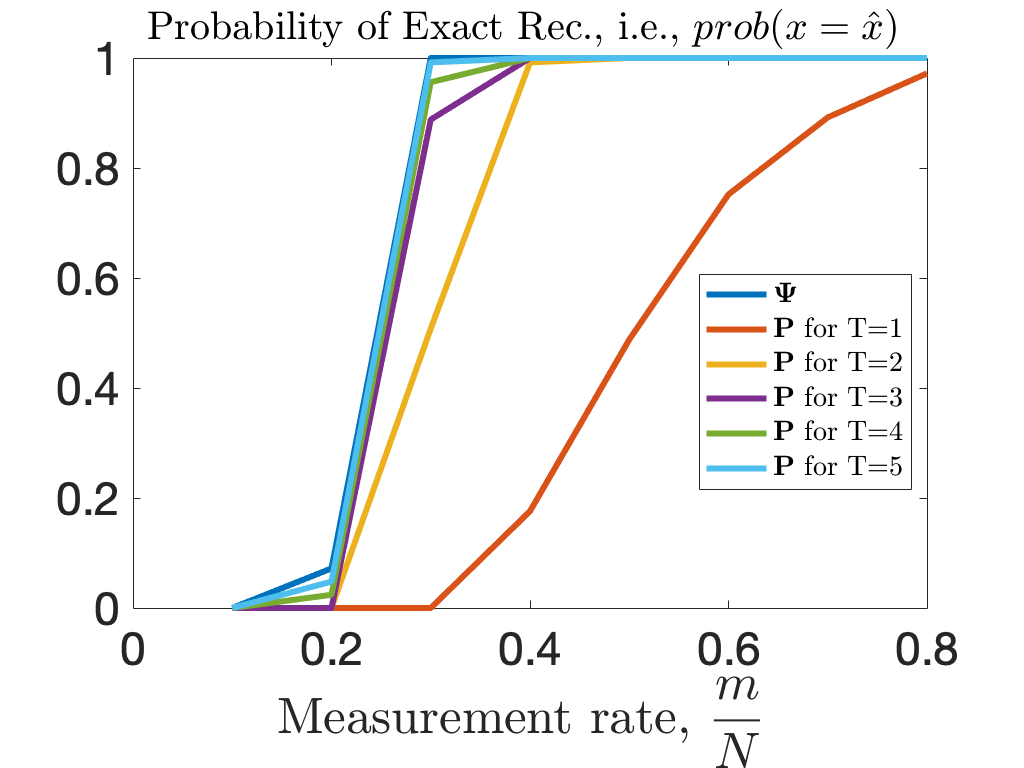}
 \caption{Average mutual coherence of the CS matrix for different realizations and calculated probability of exact recovery over 250 trials.  An exactly sparse signal is synthetically produced for $N=1024$ and $k=80$.}
 \label{mutualcoherence}
 \end{figure}

Earlier, we discussed in detail the trade-off for computational complexity versus the minimum number of measurements when we move from conventional non-separable CS scheme to separable CS setup. In the sequel, to our knowledge for the first time in literature, we will demonstrate that non-separable or unconstrained CS matrix can be approximated with the summation of tensorial sum operation. By doing this, while preserving the ``goodness" of the CS matrix (i.e., the incoherence of the matrix) as possible as close to unconstrained CS case, we can reduce the number of parameters to represent the CS matrix. This will enable us to have a feasible number of learnable parameters when we will attempt to jointly learn CS operation and recovery system using a neural network architecture. Mathematically speaking, let us sum $T$ number of different separable CS tensor obtained from $\mathcal{S}$:   
\begin{equation}
    \mathcal{Y} = \sum_{t=1}^{T} \mathbf{\mathcal{S}} \times_1 \mathbf{\Psi_1^{(t)}} \times_2 \mathbf{\Psi_2^{(t)}} ... \mathbf{\Psi_{J-1}^{(t)}} \times_J \mathbf{\Psi_J^{(t)}}, \label{tensorsum1}
\end{equation}
where $\mathbf{\Psi_i^{(t)}}$ is the $i^{th}$ dimension CS matrix of $t^{th}$ operation. Eq.~\eqref{tensorsum1} can be re-formulated in a non-separable CS setup via
\begin{equation}
     \mathbf{y} = \sum_{t=1}^{T}  \left ( \mathbf{\Psi_1^{(t)}} \otimes \mathbf{\Psi_2^{(t)}} \otimes ...\otimes \mathbf{\Psi_J^{(t)}} \right ) \mathbf{s} = 
 \sum_{t=1}^{T} \mathbf{P}^{(t)} \mathbf{s }  =   \mathbf{P} \mathbf{s}, \label{tensorsum2}
\end{equation}
where $\mathbf{P}^{(t)} = \left ( \mathbf{\Psi_1^{(t)}} \otimes \mathbf{\Psi_2^{(t)}} \otimes ...\otimes \mathbf{\Psi_J^{(t)}} \right )$ and $\mathbf{P} = \sum_{t=1}^{T} \mathbf{P}^{(t)}$. Note that Eq.~\eqref{tensorsum2} is nothing but conventional non-separable CS operation with measurement matrix $\mathbf{P}$. For special case where $T=1$, Eq.~\eqref{tensorsum1} reduces to a separable CS scheme as in Eq.~\eqref{MDCS}. Compared to the conventional CS, $\mathbf{y} = \mathbf{\Psi} \mathbf{s}$ with unconstrained CS matrix $\mathbf{\Psi}$, the number of parameters to represent the CS matrix is reduced from $mN = \prod _{j=1}^{J} m_J n_J$ to $T \sum_{j=1}^{J} m_j n_j$. 
We design an experiment to show how the goodness of the new CS matrix $\mathbf{P}$ is increased with $T$. For the goodness metric, we selected the mutual coherence and the probability of exact recovery of the $k$-sparse signal in our experimental results. As the CS matrices, we selected Gaussian random projection matrices; in the case of $\mathbf{\Psi}$ which is unconstrained, $\prod_{j=1}^{J} m_J \times \prod_{j=1}^{J} n_J$ size Gaussian matrix is produced with each element of it is randomly drawn from the Gaussian distribution. For the new case, CS matrix $\mathbf{P}$ is generated with the summation of Kronecker products of the separable random Gaussian matrices as shown in Eq.~\eqref{tensorsum2}. For this experiment, we assume that the signal of interest is $k$-sparse in Canonical domain i.e., $\mathbf{s} = \mathbf{x}$. We used a mutual coherence functional, which is closely related but slightly different version of the one introduced in previous section:  $ \mu \left ( \mathbf{A} \right ) = \max_{1\leq i \leq j \leq N} \left ( \frac{\left | <\mathbf{a_i}, \mathbf{a_j}> \right |}{ \left \| \mathbf{a_i} \right \| \left \| \mathbf{a_j} \right \|} \right )$ where $\mathbf{a_i}$ is the $i^{th}$ column of matrix $\mathbf{A}$. Figure~\ref{mutualcoherence} shows us that when $T$ increases the mutual coherence of $\mathbf{P}$ decreases as expected. We also reported the probability of exact recovery from $\mathbf{y} = \mathbf{Px}$ with different $T$ values and $\mathbf{y} = \mathbf{\Psi x}$. The sparse signal length is set to $1024$ and orthogonal matching pursuit algorithm \cite{OMP} was used as the CS recovery algorithm. Even though the lowest mutual coherence of $\mathbf{P}$ is still not as low as that of $\mathbf{\Psi}$, for $T =5$, $\mathbf{P}$ can achieve similar performance in recovery when $k = 80$. Average mutual coherence is averaged and the exact recovery probability is estimated over 250 trials. 

\subsection{Structural Tensor Sum or Transformation Basis as a part of the CS matrix}
In the literature, adjusting the measurement matrix as the multiplication of two or more matrices is a common practice. For instance, in \cite{do2011fast}, structural compressive sensing matrices are constructed as the multiplication of random permutation, an orthonormal basis, and sub-sampling matrices. Thanks to such pseudo-random matrices, faster recovery can be possible compared to the CS system with full random matrices. Moreover, in \cite{encryption2, bi-level}, CS matrix is in the form of multiplication of a sparsifying basis and a random matrix (i.e., an ordinary CS sensing matrix such as Gaussian projection matrix) i.e., $\mathbf{\Psi} = \mathbf{\Psi}^* \mathbf{\Omega'} $, where $\mathbf{\Omega'}$ is transformation domain basis, and $\mathbf{\Psi}^*$ is ordinary random CS matrix. Even though this CS system was originally proposed for CS-based encryption in the frequency domain, in Section~\ref{sec:Discussion} we will discuss that the learned CS systems in the frequency domain may slightly carry more high-frequency details compared to the learned CS system in the spatial domain. If the transformation basis is also separable like DCT, such a system can also be injected in the proposed tensorial and sum of tensorial CS scheme, i.e., $\mathbf{\Psi_i^{(t)}} = \mathbf{\Psi_i^{(t)}}^* \mathbf{\Omega_i^{(t)'}} $, where $\mathbf{\Omega_i^{(t)'}}$ is $i^{th}$-coordinate matrix of the separable transformation basis.  

\subsection{Generalized Tensor Summation Compressive Sensing Network (GTSNET)}
In this section, we propose a neural network architecture that jointly learns the CS sensing mechanism (CS matrix), and the reconstruction of the signal. The proposed network is composed of three parts: i) A CS operation, ii) Adjoint of the CS operation (or coarse estimation of the signal), and iii) a refinement module. 

\begin{figure}
    \centering
    \includegraphics[width=\columnwidth]{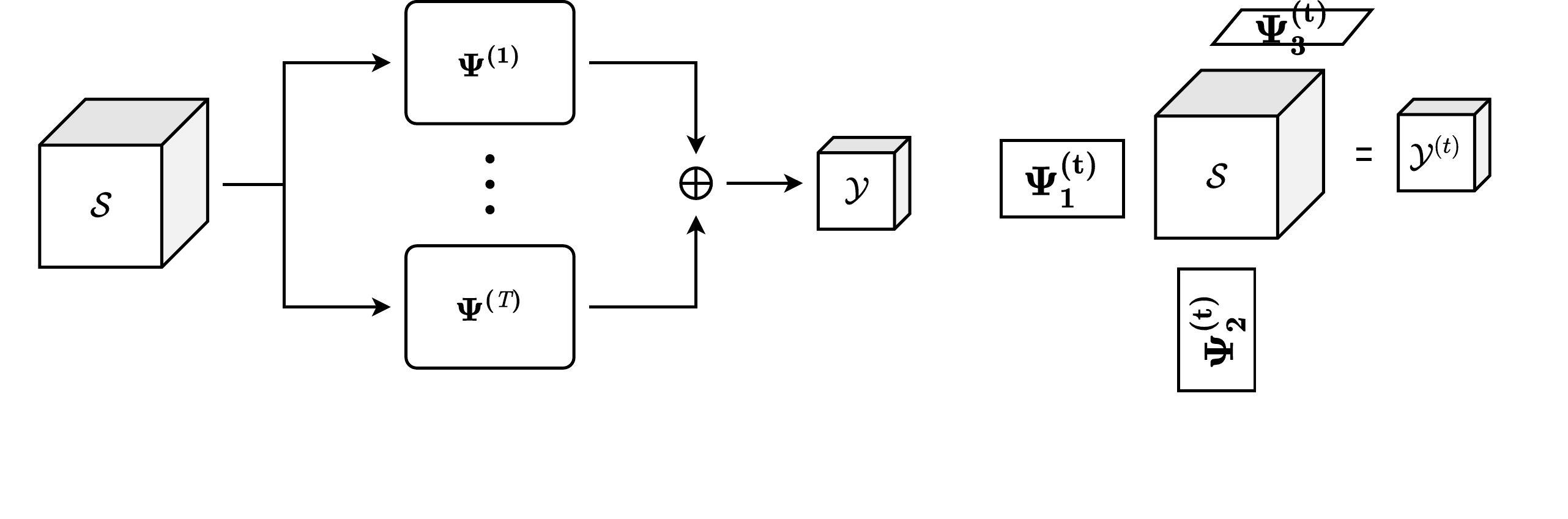}
    \caption{CS Matrix configuration. Left: The CS operation modeled as the summation of tensor sums. Right: An individual tensor sum for the case of 3D tensors.}
    \label{fig:CS_Matrix}
\end{figure}

\subsubsection{Separable and Multi-Linear Learning of CS Operation (i.e., learnable CS matrix)}
\label{subsubsec:LearnableCS}
Our learnable CS matrix $\mathbf{P}$ is factorized as
\begin{equation}
    \mathbf{P} = \sum_{t=1}^{T}  \left ( \mathbf{\Psi_1^{(t)}} \otimes \mathbf{\Psi_2^{(t)}} \otimes ...\otimes \mathbf{\Psi_J^{(t)}} \right ), \label{factorization of P}
\end{equation}
where $\mathbf{\Psi_i^{(t)}} = \mathbf{\Psi_i^{(t)}}^* \mathbf{{\Omega_i^{(t)'}}} $, $\mathbf{{\Omega_i^{(t)'}}} $ is $i^{th}$-coordinate matrix of the $t^{th}$ separable transformation matrix such as the one that represents $8x8$ size block-wise 2D DCT on the horizontal axis and $\mathbf{\Psi_i^{(t)}}^*$ learnable $i^{th}$-coordinate matrix of the $t^{th}$ term in the summation. The CS operation can be factorized using a reasonable number of training parameters thanks to the mode-j product: 

\begin{equation}
    \mathcal{Y} = \sum_{t=1}^{T} \mathbf{\mathcal{S}} \times_1 \left ( \mathbf{\Psi_1^{(t)}}^* \mathbf{{\Omega_1^{(t)'}}} \right ) \times_2 ...  \times_J \left ( \mathbf{\Psi_J^{(t)}}^* \mathbf{{\Omega_J^{(t)'}}} \right ). \label{GTSNETCS}
\end{equation}
A schematic explanation of Eq.~\eqref{GTSNETCS} is given in Figure~\ref{fig:CS_Matrix}.
\begin{itemize}
    \item For special case, $\mathbf{{\Omega_1^{(t)}}} = \mathbf{I}$, the CS system reduces to an un-structural tensor sum system, which is the learned version of the CS system defined in Eq.~\eqref{tensorsum1}.
    \item For $T=1$, the unconstrained system reduces to a separable CS system (e.g., the learned version of the separable CS imaging \cite{separableCS}).
    \item The system is valid for compressively sensing of any signal, $\mathcal{S}$.
    \item Thanks to the formulation in Eq.~\eqref{GTSNETCS}, the number of trainable parameters for unconstrained CS matrix is reduced from $\prod _{j=1}^{J} m_J n_J$ to $T \sum_{j=1}^{J} m_j n_j$, compared to conventional matrix-vector multiplication formula. Therefore, it makes the learning unconstrained CS sensing possible for large-scale and multi-dimensional signals. 
\end{itemize}
Considering these properties, we call our learnable CS operation Generalized Tensor Summation Compressive Sensing.

\begin{figure}
    \centering
    \includegraphics[width=\columnwidth]{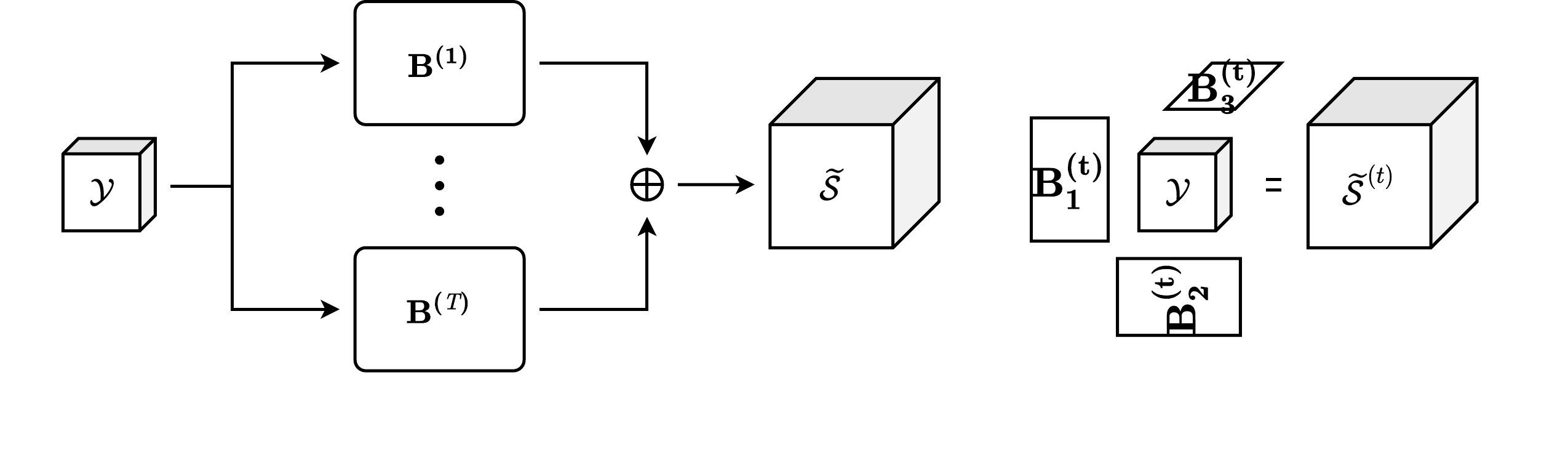}
    \caption{Course estimation module. Left: The adjoint operation modeled as the summation of tensor sums. Right: An individual tensor sum with an input tensor in 3D.}
    \label{fig:Adj_Matrix}
\end{figure}

\subsubsection{A coarse estimation of the signal: Separable Learning of adjoint of CS operation}

In traditional iterative CS reconstruction algorithms, the transposition or the pseudo-inverse of the CS matrix is used in each iteration. On the other hand, in reconstruction-free inference tasks over CS signals \cite{degerli,inference} or non-iterative deep learning-based recovery algorithms \cite{reconnet}, a coarse estimation of the signal, also known as the proxy of the signal is first obtained, i.e., $\mathbf{\widetilde{s}} = \mathbf{\Psi'} \mathbf{y}$. Although, it is also possible to obtain such a proxy using the regularized pseudo inverse of CS matrix, i.e., $ \mathbf{\widetilde{s}} = \left ( \mathbf{\Psi}' \mathbf{\Psi} + \lambda \mathbf{I} \right )^{-1} \mathbf{\Psi}' \mathbf{y}$ \cite{CSEN}, we follow the notation with the transpose or adjoint operator in general for simplicity. Eventually, the adjoint operator will be a learnable linear transformation. The adjoint of the $\mathbf{P}$ that is defined with the factorization in Eq.~\eqref{factorization of P} can simply be expressed as,
\begin{align}
 \begin{split}
    \mathbf{P'} &= \sum_{t=1}^{T}  \left ( \mathbf{\Psi_1^{(t)}} \otimes \mathbf{\Psi_2^{(t)}} \otimes ...\otimes \mathbf{\Psi_J^{(t)}} \right )' \\ &= \sum_{t=1}^{T}  \left ( \mathbf{{\Psi_J^{(t)}}'} \otimes \mathbf{{\Psi_{J-1}^{(t)}}'} \otimes ...\otimes \mathbf{{\Psi_1^{(t)}}'} \right ). \label{factorization of P transpose}
\end{split}
\end{align}
where $\mathbf{{\Psi_i^{(t)}}'}  = \mathbf{{\Omega_i^{(t)}}} \mathbf{\left ( {{\Psi_i^{(t)}}}^*  \right )'}$, $\mathbf{{\Omega_i^{(t)}}} $ is $i^{th}$-coordinate matrix of the inverse of the $t^{th}$ separable transformation matrix such as the one that represents the inverse operation of the $8x8$ size block-wise 2D DCT transformation on the horizontal axis and $\mathbf{\left ( {{\Psi_i^{(t)}}}^*  \right )'}$ is the transpose of $\mathbf{\left ( {{\Psi_i^{(t)}}}^*  \right )}$. As stated above we introduce to learn the adjoint CS matrix from the training set. Mathematically speaking, we wish to have the proxy signal, $\mathbf{\widetilde{s}} = \mathbf{B} \mathbf{y}$ where the operation $\mathbf{B}$ is learned by a neural network instead of directly applying $\mathbf{P'}$. In practice, there is no need to formulate it in vector-matrix multiplication formulation since the adjoint can be applied directly on the tensorized measurement. As shown in Figure~\ref{fig:Adj_Matrix}, it can be expressed as,
\begin{equation}
    \mathcal{\widetilde{S}} = \sum_{t=1}^{T} \mathbf{\mathcal{Y}} \times_1 \left (  \mathbf{{\Omega_1^{(t)}}} \mathbf{B_1^{(t)}}^*\right ) \times_2 ...  \times_J \left (  \mathbf{{\Omega_J^{(t)}}} \mathbf{B_J^{(t)}}^*\right ), \label{LearnableAdjoint}
\end{equation}
where $\mathbf{B_i^{(t)}}^*$ is the $i^{th}$- coordinate learnable adjoint operation matrix for the $t^{th}$ term and $\mathbf{\Omega_i^{(t)}}$ is the corresponding fix (non-trainable) inverse transformation operation. As it was in the case of CS operation, the tensorial factorization in Eq.~\eqref{LearnableAdjoint} makes the adjoint operation trainable instead of attempting directly to learn the elements of the unconstrained matrix $\mathbf{B}$. 

\begin{figure*}
    \centering
    \includegraphics[width=\textwidth]{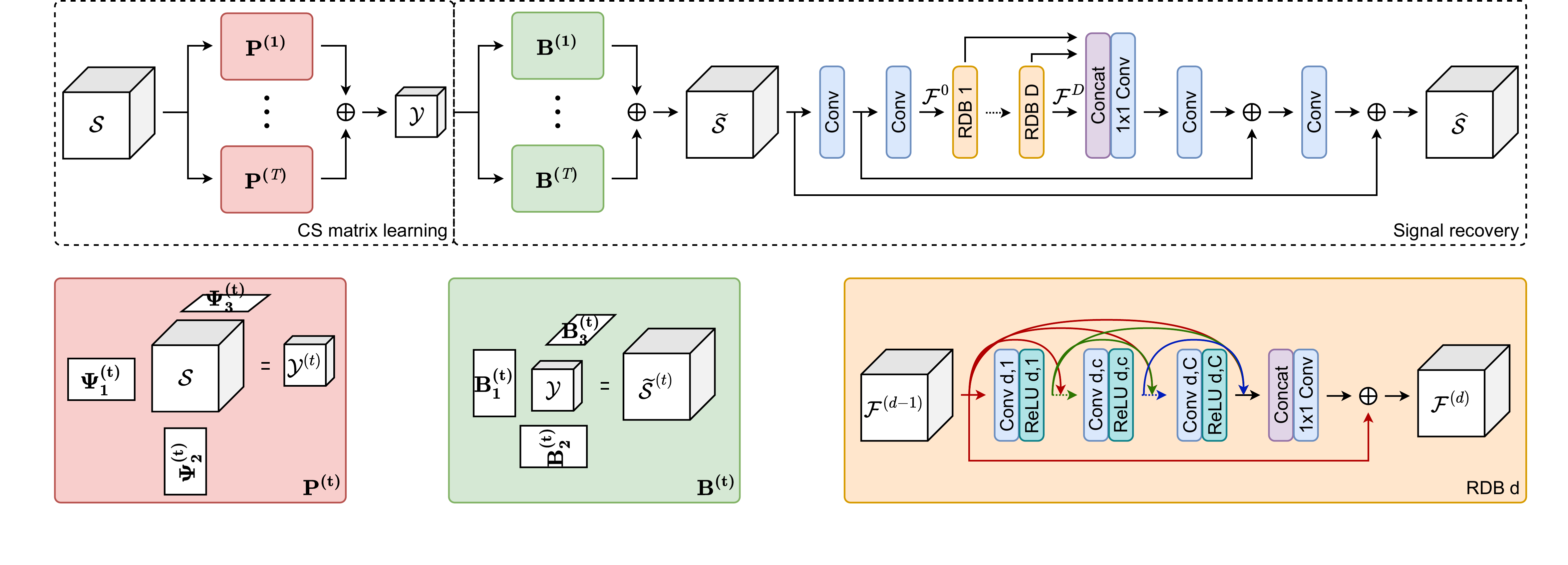}
    \caption{Overall block diagram of the proposed end-to-end system.}
    \label{fig:GTSNET_Overall}
\end{figure*}

%

\subsubsection{Reconstruction Free Recovery with Deep Neural Network}
Having the proxy signal, $\mathcal{\widetilde{S}}$, we train a conventional CNN, $C(.)$, which takes the proxy signal, $\mathcal{\widetilde{S}}$ as input and produce a finer estimation of the signal, i.e., $\mathcal{\widehat{S}} = C(\mathcal{\widetilde{S}})$. In that way, a non-iterative reconstruction network and CS operation can jointly be optimized (learned). The final network is called Generalized Tensor Summation Compressive Sensing Network (GTSNET-T) which includes $T$ tensor summation in its formula. Our solution introduces a generalized but flexible learning paradigm, it encapsulated many special cases which are set as the adjustable parameter of the network. For instance, one can set $T=1$ to have a separable optimal CS operation and its reconstruction. Or alternatively, the transformation $\mathbf{{\Omega_i^{(t)}}} $ can be set to the identity operator to sense in the spatial domain.

For the refinement module, $C(.)$, we incorporate a modified version of the Residual Dense Network (RDN) \cite{RDN}. Such network takes advantage of the so-called residual dense blocks (RDBs), within which all the layer outputs are fully utilized via local feature fusion. The outputs of each RDB are further connected via a global feature fusion, where the information from each block is effectively preserved. The performance of the network is improved by both local and global residual learning. We modify the original RDN configuration by omitting the upscale layer as it was proposed for image super-resolution \cite{RDN}. Furthermore, we adapt the overall RDN structure as a residual network, i.e., the output of the modified RDN is added to the input proxy signal, $\mathcal{\widetilde{S}}$, to obtain $\mathcal{\widehat{S}}$.     

The overall GTSNET-T structure is illustrated in Figure~\ref{fig:GTSNET_Overall}, including the learnable CS matrix and the adjoint operation, as well as the final refinement module. The adjoint operation matrices $\mathbf{{\Omega_j^{(t)}}} \mathbf{B_j^{(t)}}^*$ are denoted altogether as $\mathbf{B_j^{(t)}}\ = \mathbf{{\Omega_j^{(t)}}} \mathbf{B_j^{(t)}}^*$ for simplicity. Each branch $t$ in the CS matrix (pink blocks in Figure~\ref{fig:GTSNET_Overall}) performs a single tensor product with the input tensor $\mathcal{S}$, while the final CS operation is the summation over the products as dictated by Eq.~\eqref{tensorsum1} and Eq.~\eqref{tensorsum2}. Similarly, the adjoint operation is the summation over the individual tensor products with the compressed signal $\mathcal{Y}$ (green blocks in Figure~\ref{fig:GTSNET_Overall}), as given by Eq.~\eqref{LearnableAdjoint}. The refinement module composes of D RDBs, each of which having C convolution layers with rectified linear units (ReLUs) \cite{Relu} as activation functions. The input feature map to each RDN, $\mathcal{F}^{(d-1)}$ has G$_0$ channels, while each convolution layer inside has G filters with a $3\times3$ filter size. The concatenated feature maps at the end of the RDB are processed through one $1\times1$ convolution layer to map the output channel size back to G$_0$. In our implementation, we set $\text{D} = 4$, $\text{C} = 3$, $\text{G}_0 = 30$, and $\text{G} = 12$. 

We train the network with an overall loss function $L(\mathcal{\widetilde{S}},\mathcal{\widehat{S}},\mathcal{S})$, which is the combination of two loss functions defined over the proxy signal, $\widetilde{S}$, and the final output, $\widehat{S}$, respectively. The loss over the proxy signal, $\widetilde{L}(\mathcal{\widetilde{S}},\mathcal{S})$, is set to be a simple $L_1$-loss, as it was previously demonstrated to achieve better performance compared to the $L_2$-loss in various image processing problems \cite{L1_loss}. Assuming each training batch contains K input tensors $\{\mathcal{S}_1,\mathcal{S}_2,...,\mathcal{S}_K\}$,
\begin{equation}
    \widetilde{L}(\mathcal{\widetilde{S}},\mathcal{S}) = \frac{1}{K} \sum_{k=1}^{K} \left \| \mathbf{\widetilde{s}}_k - \mathbf{s}_k \right \|_1. 
\end{equation}
The loss over the final output, $\widehat{L}(\mathcal{\widehat{S}},\mathcal{S})$, is also an $L_1$-loss with an additional regularization term, $\widehat{R}(\mathcal{\widehat{S}},\mathcal{S})$,
\begin{equation}
    \widehat{L}(\mathcal{\widehat{S}},\mathcal{S}) = \frac{1}{K}  \sum_{k=1}^{K} \big( \left \| \mathbf{\widehat{s}}_k - \mathbf{s}_k \right \|_1 +\alpha \widehat{R}(\mathcal{\widehat{S}}_k,\mathcal{S}_k) \big), 
\end{equation} 
where $\alpha$ is a hyperparameter. Our regularization term is a modified sparse gradient prior \cite{GradientPrior} applied on the spatial domain, which has been proposed for image debluring as it provides sharper details compared to, e.g., Gaussian prior. The mathematical description of the regularization is expressed as follows:
\begin{align}
 \begin{split}
    \widehat{R}(\mathcal{\widehat{S}},\mathcal{S}) = \sum_{n_1,...,n_J} \exp{(-\beta |\nabla_{n_1} \mathcal{S}|^\gamma)}  |\nabla_{n_1} \mathcal{\widehat{S}}|^\gamma + \\ \sum_{n_1,...,n_J} \exp{(-\beta |\nabla_{n_2} \mathcal{S}|^\gamma)}  |\nabla_{n_2} \mathcal{\widehat{S}}|^\gamma,
\end{split}
\label{eq:Regularization}
\end{align}
where $\nabla_{n_1}$ and $\nabla_{n_2}$ are the discrete differential operators over the first and second dimensions, respectively. The exponential weights $\exp{(-\beta |\nabla_{n_1} \mathcal{S}|^\gamma)}$ and $\exp{(-\beta |\nabla_{n_2} \mathcal{S}|^\gamma)}$ are introduced to decrease the prior term over the edges of the original tensor $\mathcal{S}$, as proposed in \cite{GradientPrior}. We empirically set, $\alpha=0.005$, $\beta=10$, and $\gamma=0.9$. Finally, the overall loss function is $L(\mathcal{\widetilde{S}},\mathcal{\widehat{S}},\mathcal{S}) = \widetilde{L}(\mathcal{\widetilde{S}},\mathcal{S}) + \widehat{L}(\mathcal{\widehat{S}},\mathcal{S})$.





\section{Experimental Setup and Results}
\label{results}
\subsection{Training Setup}
We prepare the training dataset in the following manner: Div2K image dataset was used and $256 \times 256$ image patches were selected with stride 512 and they were cropped. Data augmentation was applied during the data generation with rotations in four different degrees; 0, 90, 180, and 270, flipping and downsampling with scale factors; 1, 0.8, and 0.6. Hence, by using the training set of DIV2K total of 89272 image patches were obtained to be used as the training set. Similarly, as the validation set, we obtained 1512 images from the validation set of DIV2K. All the images are normalized to range $\left [0,1  \right ]$. The batch size was selected as 16 and the networks were trained with 100 epochs. During training, the learning rates are scheduled to be $10^{-3}$ for the first 50 epoch, $10^{-4}$ for the later 30 epoch, and $10^{-5}$ for the last 20 epoch. The network of the 100. epoch was chosen as final. The implementation of the GTSNET was done using MatConvNet package \cite{Matconvnet}. 
\subsection{Comparative Evaluations}

\subfile{results/table_compare_soa_gray}

\subfile{results/figure_compare_soa_gray}

As traditional CS reconstruction methods, which are well-known state-of-the-art sparse recovery methods, comparative evaluations are performed against the following three methods; Gradient Projection for Sparse Reconstruction (GPSR) \cite{figueiredo2007gradient}, TV Minimization by Augmented Lagrangian and Alternating Direction Algorithms (TVAL3) \cite{TVAL3} and Denoising-based AMP (D-AMP) \cite{DAMP}. 
GPSR is a sparse recovery algorithm that was specifically proposed as computationally more efficient and feasible to apply for any image CS framework. As the CS matrix, a randomly selected subset of the rows of noiselet basis \cite{imaging} was used. As the sparsifying transform, wavelet "Coiflet 2" was used with the toolbox WaveLab850 \cite{donoho2006wavelab}.  TVAL3 is one of the state-of-the-art TV minimization solvers.  Walsh Hadamard Transform whose fast implementation available in the TVAL3 toolbox was used as the CS matrix. The parameters on TVAL3 toolbox were set as follows: $\mu = 2^{13}, \beta = 2^6, \mu_0 = 2^2, \beta_0 =  2^{-2}, tol=10^{-6}, maxit =300$. D-AMP was proposed to improve the performance of CS recovery for the natural signals by employing off-the-shelf denoising algorithms. We test the algorithm with default settings, where the elements of the CS matrix are picked from \textit{i.i.d.} Gaussian distribution and BM3D \cite{BM3D} is used as the denoiser. The number of iterations and the image block size are empirically set as 30 and $128\times 128$, respectively.

As the akin state-of-the-art deep learning methods, we selected CSNET \cite{CSNET} and SCSNET \cite{SCSNET}. The algorithms and the trained models were taken from the competing algorithms' web pages. Both methods jointly learn the CS matrix and reconstruction of the image from the measurement as proposed in this study. However, these methods learn the block-wise ($96 \time 96$) CS matrix using convolution operation in a non-overlapping manner. In that sense, when the kernel size is increased to full image size, the method turns out to be the classical unconstrained CS setup with an infeasible increase in the number of parameters to train.   

We trained two GTSNET versions; GTSNET-1 and GTSNET-3. Among them GTSNET-1 learns tensorial representation of CS matrix, therefore suitable for both separable and unconstrained CS schemes. For this network, separable transformation matrices  $\Omega_1'$ and $\Omega_2'$ were chosen as $8 \times 8$ DCT transformation matrices in the horizontal and vertical directions, respectively.  GTSNET-3 includes the three-tensor summation as the CS operation and represents an unconstrained CS setup. As the sparsifying matrices, we selected, $8 \times 8$, $16 \times 16$ and $32 \times 32$ 2D DCT transformations for $\Omega^{(1)'}$, $\Omega^{(2)'}$ and $\Omega^{(3)'}$. All the competing algorithms were tested on three commonly-used datasets: Set14 \cite{Set14}, Set5 \cite{SET5}, and Set11 \cite{reconnet}. The results on five different measurement rates (MRs) are presented in Table \ref{tab:sotagray}. Against the competing traditional methods, GPRS, TVAL3 and DAMP, a significant gap on the average performance is observed. In particular, we achieve 7.21 dB, 4.55 dB, 1.6 dB, 1.63 dB, and 2.81 dB improvements in PSNRs compared to the closest performance, for MRs of 0.01, 0.05, 0.1, 0.2 and 0.3, respectively. When we compare against the deep learning-based competing methods, CSNET+ and SCSNET, GTSNET-T shows superiority for the lower MRs ($<$ 0.2), and achieves a comparable performance for higher MRs, i.e., 0.3 dB, 0.3 dB, 0.18 dB PSNR improvement over the best competing method, for the MRs of 0.01, 0.05 and 0.1, respectively. Figure~\ref{tab:sotagray} presents visual comparisons over the state-of-the-art CS methods. Although there is no significant gap between the PSNR and SSIM of GTSNET-1 and GTSNET-3 results, one can observe GTSNET-3 outputs preserve high frequency details better, e.g., see Parrot and Flinstone images in Figure~\ref{figure:sotagray} and Figure~\ref{figure:deepmethods}. The performance gap in both PSNR and SSIM measures becomes significant in RGB images while the visual quality of the GTSNET-3 outputs especially at the fine details noticeably improves.


\subsection{Comparative evaluations against deep learning-based CS methods}

\subfile{results/figure_compare_soa_DL_gray}

As the competing deep learning-based solutions, \begin{enumerate*}[label=(\roman*)]  
\item the stacked denoising autoencoder (SDA) \cite{SDA}, which is the pioneer method,
\item non-iterative reconstruction of the compressively sensed images using CNN (ReconNet) \cite{reconnet}
\item the learned version of iterative shrinkage thresholding algorithm for CS imaging (ISTA-Net),
\item akin state of the art techniques convolutional compressive sensing network (CSNET) \cite{CSNET}
\item scalable convolutional compressive sensing network (SCSNET) \cite{SCSNET},
\item memory augmented cascading Network (MAC-Net) \cite{MAC-Net} and
\item dual-path attention network for compressed sensing (DPA-Net) \cite{DPA-Net} 
\end{enumerate*}
as the most recent techniques are selected. For ISTA-Net and CSNET, we choose their improved versions ISTA-Net+ and CSNET+, respectively. The comparative evaluations are conducted on the benchmark SET11 dataset. The results for different measurement rates are presented in Table \ref{tab:deepcompare}. All the algorithms and the trained models were downloaded from authors' web pages and run over SET11 except SDA and DPA-Net, whose source codes are not available online. The results of SDA were taken from \cite{reconnet} and the results of DPA-Net were taken from \cite{DPA-Net}. The average PSNR and SSIM values show the superiority of the proposed network over all competing methods especially for the case of lower sampling rates, e.g., for $\text{MR} < 0.25$. Figure~\ref{figure:deepmethods} shows samples for the qualitative performance comparison where it is clear that the outputs of SDA, ReconNet, and ISTA-Net+ may exhibit strong blocking artifacts. The reason is that they use block-by-block sampling strategy to compressively sense the signal, and then apply block-by-block recovery strategy. On the other hand, CSNET, MAC-Net, and SCSNet algorithms have block-by-block compressive sensing setup, but their reconstruction step recovers the image as a whole by using convolutional layers. Therefore, their outputs show fewer blocking artifacts. On the other hand, GTSNET-1 CS module is convenient for both separable and unconstraint (conventional vector-matrix CS system) CS setup. When it comes to reconstruction, it uses a CNN similar to CSNET and SCSNET and recovers the image as a whole. For the use case, where one wants to use a traditional sampling setup with a better approximation of unconstraint CS matrices, the GTSNET-T ($T >1$) can be used. The sampling strategies of deep learning methods are summarized in Table \ref{tab:sampling strategies}. Although there is no significant gap in PSNR and SSIM values on average, GTSNET-3 can recover more high-frequency details as seen in the Parrot image in Figure~\ref{figure:deepmethods}. In Section~\ref{sec:Discussion}, we will discuss the effects of the tensor sum in the frequency domain.

\begin{table}[h]
\caption{Sampling and recovery strategies of the deep learning-based algorithms. GTSNET-T can be used for both classical (unconstraint) CS and separable CS systems. }
\label{tab:sampling strategies}
\centering
\begin{tabular}{|l|l|l|}
\hline
\textcolor{red}{\textbf{Algorithm}}                                                                                  & \multicolumn{1}{c|}{\begin{tabular}[c]{@{}c@{}}\textcolor{red}{\textbf{Sampling}}\\ \textcolor{red}{\textbf{Strategy}}\end{tabular}} & \multicolumn{1}{c|}{\begin{tabular}[c]{@{}c@{}}\textcolor{red}{\textbf{Reconstruction}}\\ \textcolor{red}{\textbf{Strategy}}\end{tabular}} \\ \hline
\textcolor{blue}{\textbf{SDA}}                                                                                         & \begin{tabular}[c]{@{}l@{}}Block-by-block\\ CS\end{tabular}                      & \begin{tabular}[c]{@{}l@{}}Block-by-block\\ Auto-encoder\end{tabular}                  \\ \hline
\textcolor{blue}{\textbf{ReconNet}}                                                                                    & \begin{tabular}[c]{@{}l@{}}Block-by-block\\ CS\end{tabular}                      & \begin{tabular}[c]{@{}l@{}}Block-by-block\\ CNN\end{tabular}                           \\ \hline
\textcolor{blue}{\textbf{ISTA-Net+}}                                                                                   & \begin{tabular}[c]{@{}l@{}}Block-by-block\\ CS\end{tabular}                      & \begin{tabular}[c]{@{}l@{}}Block-by-block\\ CNN\end{tabular}                           \\ \hline
\textcolor{blue}{\textbf{MAC-Net}}                                                                                         & \begin{tabular}[c]{@{}l@{}}Block-by-block\\ CS\end{tabular}                      & CNN                                                                                    \\ \hline
\textcolor{blue}{\textbf{CSNET+}}                                                                                      & \multicolumn{1}{c|}{\begin{tabular}[c]{@{}c@{}}Block-by-block\\ CS\end{tabular}} & CNN                                                                                    \\ \hline
\textcolor{blue}{\textbf{SCSNET}}                                                                                      & \multicolumn{1}{c|}{\begin{tabular}[c]{@{}c@{}}Block-by-block\\ CS\end{tabular}} & CNN                                                                                    \\ \hline
\textcolor{blue}{\textbf{GTSNET-1}}                                                                                    & \begin{tabular}[c]{@{}l@{}}Unconstraint\\ or separable\\ CS\end{tabular}         & CNN                                                                                    \\ \hline
\multicolumn{1}{|c|}{\begin{tabular}[c]{@{}c@{}}\textcolor{blue}{\textbf{GTSNET-T}}\\ ($T>1$)\end{tabular}} & \begin{tabular}[c]{@{}l@{}}Unconstraint\\ CS\end{tabular}                        & CNN                                                                                    \\ \hline
\end{tabular}
\end{table}

\begin{table}[h]
\centering
\caption{PSNR levels obtained by the competing and proposed methods over Set11 dataset.}
\begin{tabular}{|l|llll|}
\hline
\multirow{2}{*}{Algorithm} & \multicolumn{4}{c|}{Measurement Rate}                                                        \\ \cline{2-5} 
                           & 0.25                       & 0.1                        & 0.04                       & 0.01  \\ \hline
SDA                        & \multicolumn{1}{l|}{25.34} & \multicolumn{1}{l|}{22.65} & \multicolumn{1}{l|}{20.12} & 17.29 \\ \hline
ReconNet                   & \multicolumn{1}{l|}{25.60} & \multicolumn{1}{l|}{24.28} & \multicolumn{1}{l|}{20.63} & 17.27 \\ \hline
ISTA-Net+                  & \multicolumn{1}{l|}{32.44} & \multicolumn{1}{l|}{26.49} & \multicolumn{1}{l|}{21.56} & 17.45 \\ \hline
DPA-Net                    & \multicolumn{1}{l|}{31.74} & \multicolumn{1}{l|}{26.99} & \multicolumn{1}{l|}{23.50} & 18.05 \\ \hline
MAC-Net                    & \multicolumn{1}{l|}{\textbf{32.91}} & \multicolumn{1}{l|}{27.68} & \multicolumn{1}{l|}{24.22} & 18.26 \\ \hline
CSNET+                     & \multicolumn{1}{c|}{-}     & \multicolumn{1}{l|}{28.34} & \multicolumn{1}{c|}{-}     & 21.02 \\ \hline
SCSNET                     & \multicolumn{1}{c|}{-}     & \multicolumn{1}{l|}{28.52} & \multicolumn{1}{c|}{-}     & 21.04 \\ \hline
GTSNET-1                   & \multicolumn{1}{l|}{32.47} & \multicolumn{1}{l|}{28.79} & \multicolumn{1}{l|}{25.44} & \textbf{21.39} \\ \hline
GTSNET-3                   & \multicolumn{1}{l|}{32.36} & \multicolumn{1}{l|}{\textbf{28.83}} & \multicolumn{1}{l|}{\textbf{25.45}} & 21.38 \\ \hline
\end{tabular}
\label{tab:deepcompare}
\end{table}

\subsection{Comparative evaluations over RGB images}

\begin{table*}[h]
\caption{Performance metrics (PSNR and SSIM) obtained by the competing and proposed methods over four benchmark RGB image datasets.}
\setlength\tabcolsep{3.0pt}
\centering
\begin{tabular}{|l|llllllll|llllllll|}
\hline
Ratio                   & \multicolumn{8}{c|}{MR = 0.01}                                                                                                                                                    & \multicolumn{8}{c|}{MR = 0.05}                                                                                                              \\ \hline
\multirow{2}{*}{\diagbox[font=\scriptsize\itshape]{Dataset}{Method}} & \multicolumn{2}{c}{CSNET+}                            & \multicolumn{2}{c}{GTSNET-1}                           & \multicolumn{2}{c}{GTSNET-3}        & \multicolumn{2}{c|}{GTSNET-5} & \multicolumn{2}{c}{CSNET+}         & \multicolumn{2}{c}{GTSNET-1}        & \multicolumn{2}{c}{GTSNET-3}        & \multicolumn{2}{c|}{GTSNET-5} \\ \cline{2-17} 
                        & \multicolumn{1}{c}{PSNR} & SSIM                       & \multicolumn{1}{c}{PSNR} & SSIM                       & PSNR  & SSIM                       & PSNR          & SSIM         & PSNR  & SSIM                       & PSNR  & SSIM                       & PSNR  & SSIM                       & PSNR          & SSIM         \\ \hline
SET5                    & \textbf{24.35}                    & \multicolumn{1}{l|}{0.858} & 23.17                    & \multicolumn{1}{l|}{0.843} & 24.23 & \multicolumn{1}{l|}{\textbf{0.865}} & 24.16         & \textbf{0.865 }       & 29.18 & \multicolumn{1}{l|}{0.940} & 28.53 & \multicolumn{1}{l|}{0.935} & 29.82 & \multicolumn{1}{l|}{0.948} & \textbf{30.48}         & \textbf{0.955}        \\
SET14                   & \textbf{22.83 }                   & \multicolumn{1}{l|}{0.734} & 21.55                    & \multicolumn{1}{l|}{0.707} & 22.62 & \multicolumn{1}{l|}{\textbf{0.736}} & 22.49         & 0.732        & 26.71 & \multicolumn{1}{l|}{0.869} & 25.43 & \multicolumn{1}{l|}{0.818} & 26.41 & \multicolumn{1}{l|}{0.849} & \textbf{27.15}         & \textbf{0.872 }       \\
Manga109                & 21.18                    & \multicolumn{1}{l|}{0.821} & 20.43                    & \multicolumn{1}{l|}{0.825} & \textbf{21.38} & \multicolumn{1}{l|}{0.842} & 21.35         & \textbf{0.843 }       & 25.41 & \multicolumn{1}{l|}{0.914} & 26.04 & \multicolumn{1}{l|}{0.929} & 27.48 & \multicolumn{1}{l|}{0.946} & \textbf{28.36 }        & \textbf{0.953}        \\
Urban100                & \textbf{20.93}                    & \multicolumn{1}{l|}{\textbf{0.697}} & 19.51                    & \multicolumn{1}{l|}{0.649} & 20.61 & \multicolumn{1}{l|}{0.692} & 20.31         & 0.686        & \textbf{25.05 }& \multicolumn{1}{l|}{0.857} & 23.01 & \multicolumn{1}{l|}{0.794} & 24.40 & \multicolumn{1}{l|}{0.843} & 24.97         & \textbf{0.858}        \\ \hline
Avg.                    & \textbf{22.32 }                   & \multicolumn{1}{l|}{0.778} & 21.17                    & \multicolumn{1}{l|}{0.756} & 22.21 & \multicolumn{1}{l|}{\textbf{0.784}} & 22.08         & 0.782        & 26.59 & \multicolumn{1}{l|}{0.895} & 25.75 & \multicolumn{1}{l|}{0.869} & 27.03 & \multicolumn{1}{l|}{0.897} & \textbf{27.74}         & \textbf{0.910}        \\ \hline
                        & \multicolumn{8}{c|}{MR = 0.1}                                                                                                                                                     & \multicolumn{8}{c|}{MR=0.2}                                                                                                                 \\ \cline{2-17} 
SET5                    & 32.07                    & \multicolumn{1}{l|}{0.966} & 31.16                    & \multicolumn{1}{l|}{0.959} & \textbf{32.97} & \multicolumn{1}{l|}{\textbf{0.971}} & 32.45         & 0.968        & 35.15 & \multicolumn{1}{l|}{0.979} & 34.13 & \multicolumn{1}{l|}{0.975} & 35.33 & \multicolumn{1}{l|}{0.981} & \textbf{37.28  }       & \textbf{0.986   }     \\
SET14                   & 29.31                    & \multicolumn{1}{l|}{\textbf{0.926}} & 27.30                    & \multicolumn{1}{l|}{0.870} & \textbf{29.35 }& \multicolumn{1}{l|}{0.922} & 28.53         & 0.898        & 32.26 & \multicolumn{1}{l|}{0.960} & 29.79 & \multicolumn{1}{l|}{0.922} & 31.35 & \multicolumn{1}{l|}{0.944} & \textbf{34.15 }        & \textbf{0.972 }       \\
Manga109                & 28.82                    & \multicolumn{1}{l|}{0.957} & 29.49                    & \multicolumn{1}{l|}{0.961} & \textbf{31.63} & \multicolumn{1}{l|}{\textbf{0.976}} & 31.04         & 0.973        & 30.96 & \multicolumn{1}{l|}{0.969} & 33.62 & \multicolumn{1}{l|}{0.983} & 34.67 & \multicolumn{1}{l|}{0.987} & \textbf{36.11}         & \textbf{0.990 }       \\
Urban100                & \textbf{27.78}                    & \multicolumn{1}{l|}{\textbf{0.918}} & 25.13                    & \multicolumn{1}{l|}{0.861} & 27.74 & \multicolumn{1}{l|}{\textbf{0.918}} & 26.88         & 0.903        & 30.64 & \multicolumn{1}{l|}{0.955} & 27.94 & \multicolumn{1}{l|}{0.920} & 29.94 & \multicolumn{1}{l|}{0.949} & \textbf{33.11}         & \textbf{0.972}        \\ \hline
Avg.                    & 29.50                    & \multicolumn{1}{l|}{0.942} & 28.27                    & \multicolumn{1}{l|}{0.913} & \textbf{30.42} & \multicolumn{1}{l|}{\textbf{0.947}} & 29.73         & 0.936        & 32.25 & \multicolumn{1}{l|}{0.966} & 31.37 & \multicolumn{1}{l|}{0.950} & 32.82 & \multicolumn{1}{l|}{0.965} & \textbf{35.16}         & \textbf{0.980 }       \\ \hline
\end{tabular}
\label{tab:rgbtable}
\end{table*}

\begin{figure*}[]
 \centering
  \includegraphics[width=0.7\linewidth]{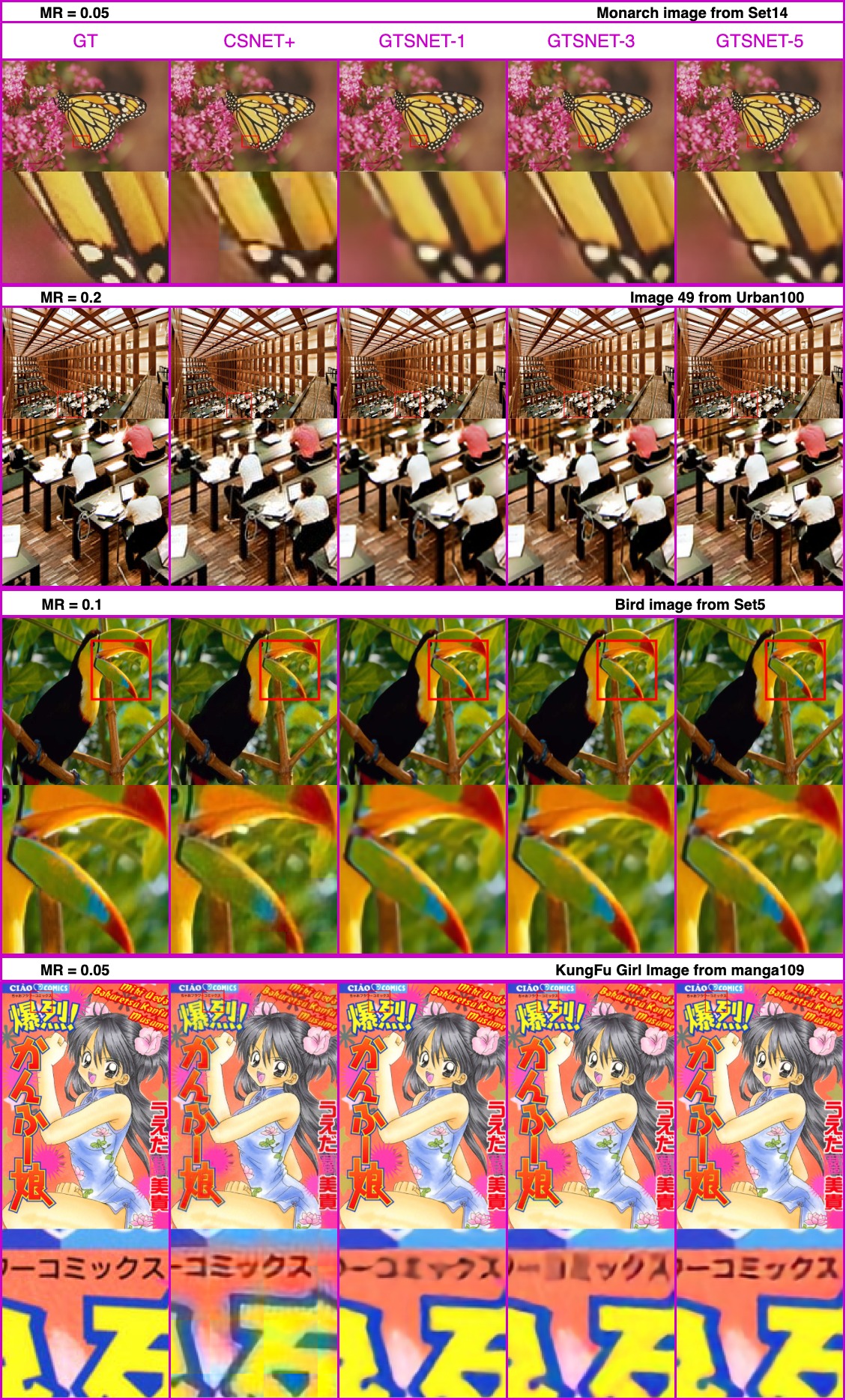}
 \caption{The recovered images of the competing and proposed methods with the GT image on the left.}
\label{fig:rgbimages}
\end{figure*}

Unfortunately, most aforementioned competing methods except CSNET were designed only for gray-scale images. Therefore, we compare GTSNET with CSNET. An extensive set of comparative evaluations was conducted on the following benchmark RGB image datasets: Set5, Set11, Manga109 \cite{manga109}, and Urban100 \cite{urban100}. The results are reported in Table \ref{tab:rgbtable}. For GTSNET-5, as the sparsifying matrices, we selected, $8 \times 8$, $16 \times 16$, $32 \times 32$, $64 \times 64$ and $128 \times 128$ 2D DCT transformations for $\Omega^{(1)'}$, $\Omega^{(2)'}$, $\Omega^{(3)'}$, $\Omega^{(4)'}$ and $\Omega^{(5)'}$, respectively. As clearly observable from the table that the performance gap between GTSNET-1 and GTSNET-T ($T>1$) widens in terms of PSNR and SSIM. Compared to CSNET+, a comparable performance with the separable CS setup (GTSNET-1) is achieved. For the unconstraint CS matrix setup ($T>1$), the performance gap between CSNET+ and the best operating GTSNET configuration becomes significant, i.e., 1.15 dB, 0.92 dB and 2.91 dB for sampling rates of 0.05, 0.1 and 0.2, respectively. Moreover, some samples for visual comparison of the recovered images are shown in Figure~\ref{fig:rgbimages}. The outputs of CSNET+ exhibit certain level of blocking artifacts that are entirely absent in any of the outputs of the proposed GTSNET-T networks.

\section{Discussion}
\label{sec:Discussion}
\subsection{Tensor vs Tensor Sum for CS matrix Learning}

\begin{figure}
    \centering
    \hspace{-12mm}
    \captionsetup[subfigure]{labelformat=empty}
    \begin{subfigure}[t]{0.4\columnwidth}
        \subfile{results/Plot_PSNRs_Set5_gray}    
    \end{subfigure}
    \hspace{7mm}
    \begin{subfigure}[t]{0.4\columnwidth}
        \subfile{results/Plot_PSNRs_Set5_rgb}    
    \end{subfigure}
    \caption{Quantitative reconstruction results of Set5 dataset, with varying number of tensor sums $T$. Left: Gray-scale (luminance) images. Right: Original color (RGB) images. The measurement rate is set to be 0.1 in both scenarios.}
    \label{fig:PSNR_Compare_Set5}
\end{figure}

In this section, we perform an ablation study concerning the effects of the number of tensor sums, $T$, over the final reconstruction quality. As a starting point, we plot the PSNR values of each image in Set5, sensed and reconstructed via three different setups, for $T=1, T=3,$ and $T=5$. We perform analysis on both gray-scale and RGB images, where the measurement rate is set as 0.1. The gray-scale images are constructed via taking the luminance channel of each image in YCbCr color space. The results are shown in Figure~\ref{fig:PSNR_Compare_Set5}. While the performance gap is negligible difference on the gray-scale images, we observe a significant performance improvement in reconstructing the RGB images as $T$ increases from 1 to 3, e.g., up to 2.74 dB PSNR improvement on the "woman" image. An interesting observation worth mentioning is that GTSNET-T with $T=3$ outperforms the one with $T=5$, both for each individual image in Set5, and for the average of each dataset presented in Table~\ref{tab:rgbtable}. This might seem at first contradictory to our derivations within the theoretical discussions, where we demonstrate in Section~\ref{sec:TensorSumCoherence} and Figure~\ref{mutualcoherence} that the mutual coherence decreases as $T$ increases. However, within such analysis each tensor is chosen to be composed of Gaussian random matrices. As in practice we further learn the CS matrices to improve performance over the random matrices, an inherent trade-off occurs, where the number of learned parameters increases linearly with increasing $T$. The experimental analysis shows that $T=3$ provides best of both worlds with a consistently superior image quality for MR=0.1.

\subfile{results/figure_freq_test_rgb}

To enrich the discussion above, we visually compare the three methods over a rather tricky color image: "zebra" from Set14. In particular, we examine the frequency responses of both the proxy signals, $\mathcal{\widetilde{S}}$, as the immediate reconstruction, and the final outputs, $\mathcal{\widehat{S}}$. The results are shown in Figure~\ref{fig:freq_tests_rgb}. The differences between each method are clearly visible over the frequency responses of the proxy signals (middle row), where the cut-off frequency of GTSNET-3 is higher compared to that of GTSNET-1 and GTSNET-5. Subsequently, the final output of GTSNET-3 can preserve the higher frequency information while providing better quantitative result in terms of PSNR value.

\subfile{results/figure_GTSNET3_Proxy_Branches}

Having shown the improvement in image quality with $T>1$, we now provide the information flowing from each branch of the adjoint operator. Figure~\ref{fig:Proxy_Branches} illustrate the tentative reconstruction results, where one of the branches ($\mathbf{B^{(3)}}$) performs the majority of the reconstruction over the lower frequency region and the residual high-frequency details are recovered through branches $\mathbf{B^{(1)}}$ and $\mathbf{B^{(2)}}$. In addition, the first and second branches carry information regarding the different regions of the spectrum; the support of $\mathbf{B^{(2)}}$ is more concentrated towards the low-frequency region, whereas the frequency response of the first branch contains higher frequencies. The proximal signal, $\mathcal{\widetilde{S}}$ (Figure~\ref{fig:Proxy_Branches}, fourth column), is the summation of each output, having a wider response than each individual branch.

\subsection{Tensor Sum vs Structural Tensor for CS matrix learning}

\subfile{results/figure_NoDCT_vs_DCT_rgb}

In Section~\ref{subsubsec:LearnableCS}, we discuss that the proposed method is suitable for designs of both structured and unstructured tensor summations, whereas we mainly demonstrate our results via the learned structured matrices. In this subsection, we compare a network trained for unstructured tensor sums, i.e., $\mathbf{{\Omega_1^{(t)}}} = \mathbf{I}$, with the previously discussed structured tensors. We train both setups for GTSNET-3, where the measurement rate is set as 0.1. A test image is picked from the Urban100 dataset for the experiment, carrying high-frequency components with fixed patterns. Figure~\ref{fig:NoDCT_vs_DCT} shows the results. While the differences are not prominent through visual inspection on the spatial domain at first, the proxy output of the structured tensor summation is observed to contain a wider frequency response. We also notice a decrease in the high-frequency region of the final output with the unstructured tensor sum, visible as a box in the middle of the frequency response (Figure~\ref{fig:NoDCT_vs_DCT}, bottom left), and a decrease of 0.43dB in PSNR. For such purposes, we proceed with the structured matrices. Nevertheless, it is important to demonstrate that the proposed method has flexibility generalizing various types of CS matrix designs. 


\section{Conclusion}
\label{conclusion}

We propose generalized tensor summation networks for fast and high-quality CS. Our framework incorporates end-to-end learning where the parameters of both the CS matrix and the signal recovery are jointly optimized. On the sensing part, the CS matrices are modeled as the summation of $T$ tensors, which has certain critical advantages. On one hand, the complexity and the number of parameters are greatly reduced thanks to the separability of tensors. By keeping $T = 1$, for instance, we can reduce the system to a Kronecker CS. On the other hand, the reduced rank of the separable systems can be addressed by increasing $T$, resulting in unconstrained CS matrices. In addition, we can design structured matrices by incorporating any separable basis into our framework, such as DCT. 

The reconstruction step of the proposed algorithm takes advantage of an adjoint operator to perform a tentative reconstruction, as well as a non-iterative, deep learning architecture as a refinement module. As we have demonstrated throughout a rigorous set of experiments, such a setup provides superior performance in both computation time and the reconstruction quality, in particular for low measurement rates, when compared with the recent methods including both traditional, iterative solutions as well as the learning-based models. We further note that even though increasing number of tensors decrease the mutual coherence of the CS matrix theoretically, in practice the increased number of parameters may decrease the optimization performance, suggesting a trade-off to be considered.


%

\appendices
\section{Derivative of Generalized Tensor Sum Operations for back propagation}

Eq.~\eqref{tensorsum1} and Eq.~\eqref{tensorsum2} provide a mathematical description of the proposed CS module. In practice, we implement a single custom Tensor layer $P^{(t)}(.)$ to perform a tensor product with the input signal, as in the pink blocks in Figure~\ref{fig:GTSNET_Overall},
\begin{equation}
    P^{(t)}(\mathcal{S}) = \mathcal{S} \times_1 \mathbf{\Psi_1^{(t)}} \times_2 \mathbf{\Psi_2^{(t)}} ... \mathbf{\Psi_{J-1}^{(t)}} \times_J \mathbf{\Psi_J^{(t)}}. 
\end{equation}
Here, with respect to the error, we derive the input derivatives of such layer, which are to be calculated during the backward pass. Let us first denote a $j-$ dimensional tensor multiplication,
\begin{equation}
    \mathcal{Y}^{(t)}_j = \mathcal{S} \times_1 \mathbf{\Psi_1^{(t)}} \times_2 \mathbf{\Psi_2^{(t)}} ... \times_j \mathbf{\Psi_{j}^{(t)}}, 
\end{equation}
where we can immediately see
\begin{equation}
    \mathcal{Y}^{(t)}_j = \mathcal{Y}^{(t)}_{j-1} \times_j \mathbf{\Psi_j^{(t)}},
    \label{eq:iterative_tensor}
\end{equation}
with $\mathcal{Y}^{(t)}_0 = \mathcal{S}$ and $P^{(t)}(\mathcal{S}) = \mathcal{Y}^{(t)}_J$. In the backward pass, we assume that the partial derivative of the error with respect to the output, $\partial L / \partial \mathcal{Y}^{(t)}_J$, is known. From Eq.~\eqref{eq:iterative_tensor} and the basics of linear algebra,
\begin{equation}
    \frac{\partial L}{\partial \mathcal{Y}^{(t)}_{J-1}} = \frac{\partial L}{\partial \mathcal{Y}^{(t)}_{J}} \times_J {\mathbf{\Psi_J^{(t)}}}',
\end{equation}
form which we iteratively go back to conclude
\begin{equation}
    \frac{\partial L}{\partial \mathcal{S}} = \frac{\partial L}{\partial \mathcal{Y}^{(t)}_{J}} \times_J {\mathbf{\Psi_J^{(t)}}}' \times_{J-1} {\mathbf{\Psi_{J-1}^{(t)}}}' ... \times_2 \mathbf{\Psi_2^{(t)}}' \times_1 \mathbf{\Psi_1^{(t)}}',
\end{equation}
which is nothing but the multiplication with the transposed tensor. 

To calculate the partial derivatives with respect to the individual matrices of the CS tensor, $\mathbf{\Psi_j^{(t)}}$, we can start by rearranging Eq.~\eqref{eq:iterative_tensor} in discrete form as
\begin{equation}
    \mathcal{Y}^{(t)}_j[n_1,...,n_j,...n_J] = \sum_{\tilde{n}_j} \Psi_j^{(t)}[n_j,\tilde{n}_j] \mathcal{Y}^{(t)}_{j-1}[n_1,...,\tilde{n}_j,...n_J].
    \label{eq:iterative_tensor_discrete}
\end{equation}
Eq.~\eqref{eq:iterative_tensor_discrete} can be converted to a simpler form as follows: Suppose an operator $mat_j(.)$ takes a $J-$dimensional input $\mathcal{Y}$ and convert it to a matrix $\mathbf{Y_j}$, by first permuting its dimensions so that the $j$th dimension of $\mathcal{Y}$ become the rows of $\mathbf{Y_j}$, and then rescaling so that all the other dimensions become the columns. Then
\begin{equation}
    \mathcal{Y}^{(t)}_j = mat_j^{-1}(\mathbf{\Psi_j^{(t)}} mat_j(\mathcal{Y}^{(t)}_{j-1})),
\end{equation}
where $mat_j^{-1}$ is the inverse of the above-mentioned $mat_j(.)$ operator. It is then straightforward to see that
\begin{equation}
    \frac{\partial L}{\partial \mathbf{\Psi_j^{(t)}}} = mat_j\bigg(\frac{\partial L}{\partial \mathcal{Y}^{(t)}_{j}}\bigg) mat_j(\mathcal{Y}^{(t)}_{j-1}))',
    \label{eq:iterative_derivative}
\end{equation}
Unfortunately, Eq.~\eqref{eq:iterative_derivative} requires a forward propagation of $\mathcal{Y}^{(t)}_{j-1}$ all the way up to $\mathcal{Y}^{(t)}_{J-1}$, which practically indicates another forward propagation before starting the backward pass. However, as these matrix multiplications at each dimension are relatively fast, such requirement does not add any noticeable overhead to the computational complexity. 

\section{Derivative of Loss function for back propagation}

Here we aim to provide the partial derivatives of error with respect to the final estimation $\mathcal{\widehat{S}}$, $\partial L / \partial \mathcal{\widehat{S}}$. As we discussed previously, our loss function is a combination of an $L_1$-loss over proxy, $\widetilde{L}(\mathcal{\widetilde{S}},\mathcal{S})$, and a regularized $L_1$-loss over the final output, $\widehat{L}(\mathcal{\widehat{S}},\mathcal{S})$. As the derivatives of $L_1$-losses are well-known, the derivation of the partial derivative of the regularization function with respect to $\mathcal{\widehat{S}}$, $\partial R / \partial \mathcal{\widehat{S}}$ will be sufficient. Furthermore, considering Eq.\ref{eq:Regularization} is separable into first and second dimensions, we can derive the derivative in 1D for simplicity, from which the extension to 2D is straightforward. If we then reformulate Eq.~\eqref{eq:Regularization} in 1D
\begin{align}
 \begin{split}
    \widehat{R}_{n_1}(\mathcal{\widehat{S}},\mathcal{S}) = \sum_{n_1,...,n_J} \exp{(-\beta |\nabla_{n_1} \mathcal{S}|^\gamma)}  |\nabla_{n_1}\mathcal{\widehat{S}}|^\gamma, 
\end{split}
\label{eq:Regularization1D}
\end{align}
where $\nabla_{n_1} \mathcal{\widehat{S}} = \mathcal{\widehat{S}}[n_1+1,n_2,...,n_J] - \mathcal{\widehat{S}}[n_1,n_2,...,n_J]$ in discrete form. Let us first denote
\begin{equation}
   \mathcal{\widehat{G}}_{n_1} = \nabla_{n_1} \mathcal{\widehat{S}}, \quad \mathcal{G}_{n_1} = |\mathcal{\widehat{G}}_{n_1}|, \quad \mathcal{W}_{n_1} = \exp{(-\beta |\nabla_{n_1}\mathcal{S}|^\gamma)}.
\end{equation}
From now on we can treat $\mathcal{W}_{n_1}$ as a constant, element-wise weighting factor, as it only depends on the label $\mathcal{S}$. Then, using Eq.~\eqref{eq:Regularization1D}, the first partial derivative is
\begin{equation}
   \frac{\partial R_{n_1}}{\partial \mathcal{G}_{n_1}} = \gamma \mathcal{W}_{n_1} \mathcal{G}_{n_1}^{(\gamma-1)}.
\end{equation}
Assuming the derivative of $|.|$ is $sgn(.)$, i.e., the sign function,
\begin{equation}
   \frac{\partial R_{n_1}}{\partial \mathcal{\widehat{G}}_{n_1}} = \gamma \mathcal{W}_{n_1} \mathcal{G}_{n_1}^{(\gamma-1)} sgn(\mathcal{\widehat{G}}_{n_1}).
   \label{eq:drdg}
\end{equation}
The partial derivative $\partial \widehat{G} / \partial \mathcal{\widehat{S}}$ is defined as follows
\begin{equation}
       \frac{\partial \mathcal{\widehat{G}}_{n_1}[n_1,...,n_J]}{\partial \mathcal{\widehat{S}}[\tilde{n}_1,...,\tilde{n}_J]} = 
    \begin{cases}
        1,  & \text{if } \tilde{n}_1=n_1-1 \And \tilde{n}_2,...,\tilde{n}_J = n_2,...,n_J\\
        -1, & \text{if } \tilde{n}_1=n_1 \And \tilde{n}_2,...,\tilde{n}_J = n_2,...,n_J\\
        0,  & \text{otherwise}
    \end{cases}
    \label{eq:dgds}
\end{equation}
Using Eq.~\eqref{eq:drdg} and Eq.~\eqref{eq:dgds}, and the chain rule, one can find in 1D
\begin{equation}
   \frac{\partial R_{n_1}}{\partial \widehat{S}} = \frac{\partial R_{n_1}}{\partial \mathcal{\widehat{G}}_{n_1}}[n_1-1,n_2,...,n_J] - \frac{\partial R_{n_1}}{\partial \mathcal{\widehat{G}}_{n_1}}[n_1,n_2,...,n_J].
   \label{eq:drds1D}
\end{equation}
By changing dimension from $n_1$ to $n_2$ and extending to 2D, we can conclude that
\begin{equation}
   \frac{\partial R}{\partial \widehat{S}} = \frac{\partial R_{n_1}}{\partial \widehat{S}} + \frac{\partial R_{n_2}}{\partial \widehat{S}}.
   \label{eq:drds2D}
\end{equation}



\ifCLASSOPTIONcaptionsoff
  \newpage
\fi



%
\bibliographystyle{IEEEtran}
\bibliography{references}

%








\end{document}

%% file: results/table_compare_soa_gray.tex
\begin{table*}[h]
\caption{Performance metrics (PSNR and SSIM) obtained by the competing and proposed methods over three benchmark datasets.}
\label{tab:sotagray}
\centering
\setlength\tabcolsep{4.5pt}
\begin{tabular}{|c|c|c|c|c|c|l|l|c|c|c|c|c|c|c|c|}
\hline
\multirow{2}{*}{\begin{tabular}[c]{@{}c@{}}Measurement\\ Rates (MRs)\end{tabular}} & \multirow{2}{*}{Datasets} & \multicolumn{2}{c|}{GPSR} & \multicolumn{2}{c|}{TVAL3} & \multicolumn{2}{c|}{DAMP}                             & \multicolumn{2}{c|}{CSNET+} & \multicolumn{2}{c|}{SCSNET} & \multicolumn{2}{c|}{GTSNET-1} & \multicolumn{2}{c|}{GTSNET-3} \\ \cline{3-16} 
                                                                                   &                           & PSNR        & SSIM        & PSNR         & SSIM        & \multicolumn{1}{c|}{PSNR} & \multicolumn{1}{c|}{SSIM} & PSNR         & SSIM         & PSNR         & SSIM         & PSNR          & SSIM          & PSNR          & SSIM          \\ \hline
\multirow{4}{*}{MR = 0.01}                                                         & SET5                      & 16.25       & 0.378       & 17.08        & 0.552       & 8.30                      & 0.066                     & 24.18        & 0.669        & 24.21        & 0.669        & 24.61         & 0.696         & \textbf{24.66}         & \textbf{0.697}         \\ \cline{2-16} 
                                                                                   & SET14                     & 16.61       & 0.349       & 16.46        & 0.474       & 7.69                      & 0.041                     & 22.93        & 0.588        & 22.97        & 0.588        & \textbf{23.08}         & \textbf{0.600}        & \textbf{23.08}         & \textbf{0.600}         \\ \cline{2-16} 
                                                                                   & SET11                     & 14.07       & 0.289       & 13.94        & 0.441       & 5.61                      & 0.024                     & 21.02        & 0.589        & 21.04        & 0.589        & \textbf{21.39}         & \textbf{0.609}         & 21.38         & \textbf{0.609}        \\ \cline{2-16} 
                                                                                   & Avg.                      & 15.64       & 0.339       & 15.83        & 0.489       & 7.20                      & 0.044                     & 22.71        & 0.615        & 22.74        & 0.615        & 23.03         & \textbf{0.635}         & \textbf{23.04}        & \textbf{0.635}        \\ \hline
\multirow{4}{*}{MR = 0.05}                                                         & SET5                      & 20.58       & 0.413       & 23.44        & 0.661       & 26.56                     & 0.766                     & 29.75        & 0.848        & 29.74        & 0.847        & \textbf{30.25}         & \textbf{0.861}         & 30.24         & \textbf{0.861}         \\ \cline{2-16} 
                                                                                   & SET14                     & 20.17       & 0.361       & 19.95        & 0.555       & 24.70                     & 0.658                     & 27.04        & 0.739        & 27.04        & 0.739        & \textbf{27.16}         & \textbf{0.747}         & 27.15         & \textbf{0.747}         \\ \cline{2-16} 
                                                                                   & SET11                     & 17.69       & 0.307       & 17.27        & 0.576       & 21.77                     & 0.684                     & 25.86        & 0.788        & 25.86        & 0.787        & \textbf{26.27}         & \textbf{0.806}         & 26.25         & \textbf{0.806}         \\ \cline{2-16} 
                                                                                   & Avg.                      & 19.48       & 0.360       & 20.22        & 0.597       & 24.34                     & 0.703                     & 27.55        & 0.792        & 27.55        & 0.791        & \textbf{27.89}         & \textbf{0.805}         & 27.88         & \textbf{0.805}         \\ \hline
\multirow{4}{*}{MR = 0.1}                                                          & SET5                      & 23.18       & 0.505       & 26.00        & 0.743       & 31.42                     & 0.872                     & 32.60        & 0.906        & 32.78        & 0.908        & 33.03         & 0.912         & \textbf{33.11}         & \textbf{0.913}         \\ \cline{2-16} 
                                                                                   & SET14                     & 22.31       & 0.438       & 20.76        & 0.620       & 28.77                     & 0.769                     & 29.24        & 0.820        & \textbf{29.32}        & \textbf{0.821}        & 29.17         & \textbf{0.821}         & 29.24         & \textbf{0.821}         \\ \cline{2-16} 
                                                                                   & SET11                     & 20.05       & 0.396       & 18.59        & 0.670       & 26.17                     & 0.852                     & 28.34        & 0.859        & 28.52        & 0.862        & 28.79         & 0.871         & \textbf{28.83}         & \textbf{0.872}         \\ \cline{2-16} 
                                                                                   & Avg.                      & 21.85       & 0.446       & 21.78        & 0.678       & 28.79                     & 0.831                     & 30.06        & 0.862        & 30.21        & 0.864        & 30.33         & 0.868         & \textbf{30.39}         & \textbf{0.869}         \\ \hline
\multirow{4}{*}{MR = 0.2}                                                          & SET5                      & 26.76       & 0.659       & 27.92        & 0.823       & 35.26                     & 0.926                     & 36.07        & 0.949        & 36.17        & 0.949        & \textbf{36.29}         & \textbf{0.950}         & 36.27         & \textbf{0.950}         \\ \cline{2-16} 
                                                                                   & SET14                     & 25.23       & 0.578       & 23.04        & 0.717       & 32.05                     & 0.851                     & 32.26        & 0.896        & \textbf{32.30}        & \textbf{0.897}        & 31.88         & 0.893         & 31.94         & 0.894         \\ \cline{2-16} 
                                                                                   & SET11                     & 23.49       & 0.561       & 20.85        & 0.789       & 27.89                     & 0.913                     & 31.67        & 0.921        & 31.83        & 0.922        & 31.81         & \textbf{0.924}         & \textbf{31.86}         & 0.923         \\ \cline{2-16} 
                                                                                   & Avg.                      & 25.16       & 0.599       & 23.94        & 0.776       & 31.73                     & 0.897                     & 33.33        & 0.922        & \textbf{33.43}        & \textbf{0.923}        & 33.33         & 0.922         & 33.36         & 0.922         \\ \hline
\multirow{4}{*}{MR = 0.3}                                                          & SET5                      & 29.55       & 0.763       & 26.27        & 0.843       & 36.86                     & 0.946                     & 38.29        & 0.965        & \textbf{38.49}        & \textbf{0.966}        & 37.95         & 0.964         & 38.00         & 0.964         \\ \cline{2-16} 
                                                                                   & SET14                     & 27.61       & 0.690       & 23.70        &  0.745           & 33.33                     & 0.894                     & 34.46        & 0.931        & \textbf{34.64}        & \textbf{0.933}        & 33.64         & 0.926         & 33.74         & 0.927         \\ \cline{2-16} 
                                                                                   & SET11                     & 26.48       & 0.692       & 23.21        & 0.854       & 27.11                     & 0.943                     & 34.32        & 0.950        & \textbf{34.66}        & \textbf{0.952}        & 34.09         & 0.951         & 33.98         & 0.950         \\ \cline{2-16} 
                                                                                   & Avg.                      & 27.88       & 0.715       & 24.39        & 0.814             & 32.43                     & 0.928                     & 35.69        & 0.949        & \textbf{35.93}        & \textbf{0.950}        & 35.23         & 0.947         & 35.24         & 0.947         \\ \hline
\end{tabular}
\end{table*}

%% file: results/figure_compare_soa_gray.tex
\begin{figure*}[h]	
	\centering
	\captionsetup[subfigure]{labelformat=empty}
    	\begin{subfigure}[t]{0.115\textwidth}
		\caption{GPSR}
		\centering
		\includegraphics[width=\columnwidth]{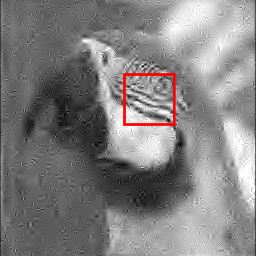}
	\end{subfigure}
	\put(-70,-98){\rotatebox{90}{MR = 0.1}}
	\vspace{-7mm}
	\begin{subfigure}[t]{0.115\textwidth}
		\caption{TVAL}
		\centering
		\includegraphics[width=\columnwidth]{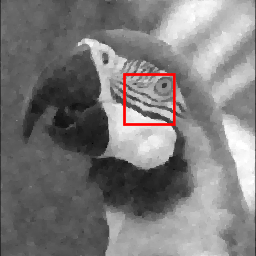}
	\end{subfigure}
	\begin{subfigure}[t]{0.115\textwidth}
		\caption{DAMP}
		\centering
		\includegraphics[width=\columnwidth]{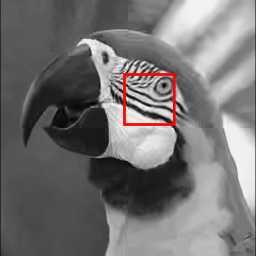}
	\end{subfigure}
	\begin{subfigure}[t]{0.115\textwidth}
		\caption{CSNet+}
		\centering
		\includegraphics[width=\columnwidth]{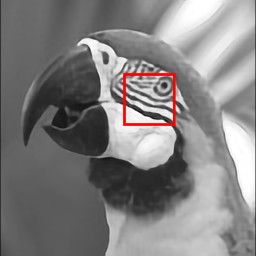}
	\end{subfigure}
	\begin{subfigure}[t]{0.115\textwidth}
		\caption{SCSNet}
		\centering
		\includegraphics[width=\columnwidth]{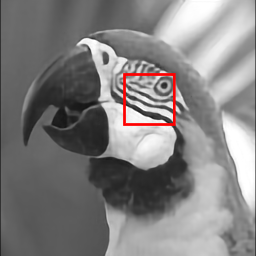}
	\end{subfigure}
	\begin{subfigure}[t]{0.115\textwidth}
		\caption{GTSNET-1}
		\centering
		\includegraphics[width=\columnwidth]{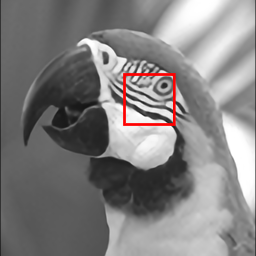}
	\end{subfigure}
	\begin{subfigure}[t]{0.115\textwidth}
		\caption{GTSNET-3}
		\centering
		\includegraphics[width=\columnwidth]{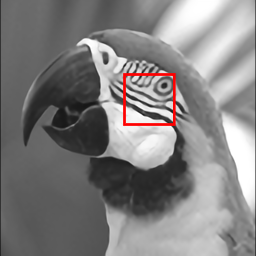}
	\end{subfigure}
	\begin{subfigure}[t]{0.115\textwidth}
		\caption{GT}
		\centering
		\includegraphics[width=\columnwidth]{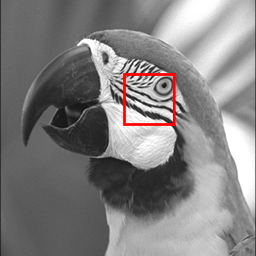}
	\end{subfigure}

	\begin{subfigure}[t]{0.115\textwidth}
		\centering
		\includegraphics[width=\columnwidth]{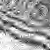}
		\caption{22.04dB}
	\end{subfigure}
	\vspace{1mm}	
	\begin{subfigure}[t]{0.115\textwidth}
		\centering
		\includegraphics[width=\columnwidth]{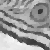}
		\caption{23.87dB}
	\end{subfigure}
	\begin{subfigure}[t]{0.115\textwidth}
		\centering
		\includegraphics[width=\columnwidth]{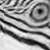}
		\caption{24.02dB}
	\end{subfigure}
	\begin{subfigure}[t]{0.115\textwidth}
		\centering
		\includegraphics[width=\columnwidth]{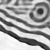}
		\caption{28.11dB}
	\end{subfigure}
	\begin{subfigure}[t]{0.115\textwidth}
		\centering
		\includegraphics[width=\columnwidth]{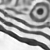}
		\caption{28.10dB}
	\end{subfigure}
	\begin{subfigure}[t]{0.115\textwidth}
		\centering
		\includegraphics[width=\columnwidth]{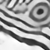}
		\caption{29.44dB}
	\end{subfigure}
	\begin{subfigure}[t]{0.115\textwidth}
		\centering
		\includegraphics[width=\columnwidth]{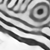}
		\caption{29.25dB}
	\end{subfigure}
	\begin{subfigure}[t]{0.115\textwidth}
		\centering
		\includegraphics[width=\columnwidth]{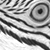}
		\caption{}
	\end{subfigure}
	
	\begin{subfigure}[t]{0.115\textwidth}
		\centering
		\includegraphics[width=\columnwidth]{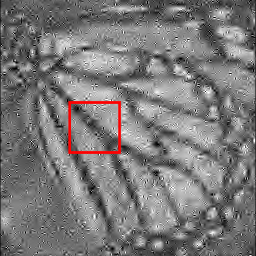}
	\end{subfigure}
    \put(-70,-23){\rotatebox{90}{MR = 0.1}}
	\vspace{-7.2mm}
	\begin{subfigure}[t]{0.115\textwidth}
		\centering
		\includegraphics[width=\columnwidth]{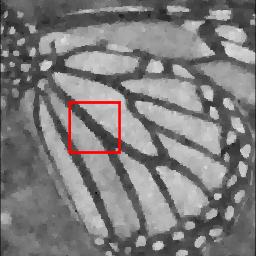}
	\end{subfigure}
	\begin{subfigure}[t]{0.115\textwidth}
		\centering
		\includegraphics[width=\columnwidth]{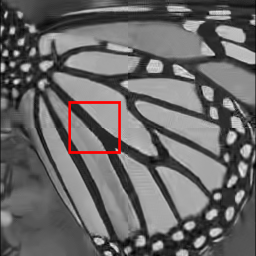}
	\end{subfigure}
	\begin{subfigure}[t]{0.115\textwidth}
		\centering
		\includegraphics[width=\columnwidth]{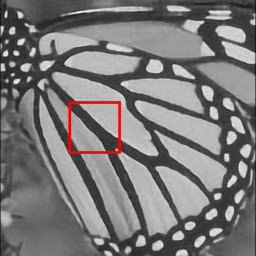}
	\end{subfigure}
	\begin{subfigure}[t]{0.115\textwidth}
		\centering
		\includegraphics[width=\columnwidth]{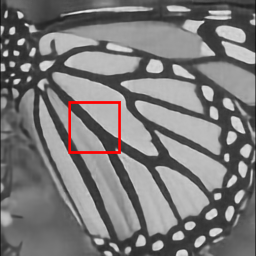}
	\end{subfigure}
	\begin{subfigure}[t]{0.115\textwidth}
		\centering
		\includegraphics[width=\columnwidth]{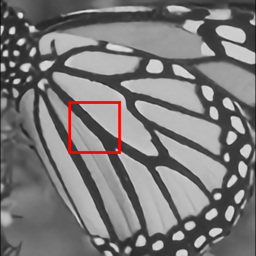}
	\end{subfigure}
	\begin{subfigure}[t]{0.115\textwidth}
		\centering
		\includegraphics[width=\columnwidth]{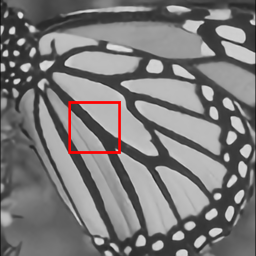}
	\end{subfigure}
	\begin{subfigure}[t]{0.115\textwidth}
		\centering
		\includegraphics[width=\columnwidth]{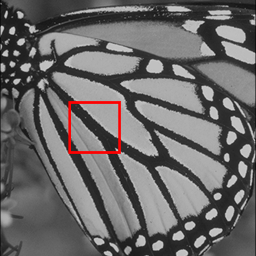}
	\end{subfigure}

	\begin{subfigure}[t]{0.115\textwidth}
		\centering
		\includegraphics[width=\columnwidth]{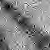}
		\caption{16.78dB}
	\end{subfigure}
	\vspace{1mm}	
	\begin{subfigure}[t]{0.115\textwidth}
		\centering
		\includegraphics[width=\columnwidth]{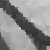}
		\caption{22.69dB}
	\end{subfigure}
	\begin{subfigure}[t]{0.115\textwidth}
		\centering
		\includegraphics[width=\columnwidth]{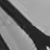}
		\caption{26.44dB}
	\end{subfigure}
	\begin{subfigure}[t]{0.115\textwidth}
		\centering
		\includegraphics[width=\columnwidth]{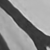}
		\caption{28.41dB}
	\end{subfigure}
	\begin{subfigure}[t]{0.115\textwidth}
		\centering
		\includegraphics[width=\columnwidth]{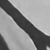}
		\caption{28.79dB}
	\end{subfigure}
	\begin{subfigure}[t]{0.115\textwidth}
		\centering
		\includegraphics[width=\columnwidth]{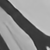}
		\caption{28.91dB}
	\end{subfigure}
	\begin{subfigure}[t]{0.115\textwidth}
		\centering
		\includegraphics[width=\columnwidth]{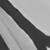}
		\caption{28.89dB}
	\end{subfigure}
	\begin{subfigure}[t]{0.115\textwidth}
		\centering
		\includegraphics[width=\columnwidth]{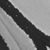}
		\caption{}
	\end{subfigure}
	
	\begin{subfigure}[t]{0.115\textwidth}
		\centering
		\includegraphics[width=\columnwidth]{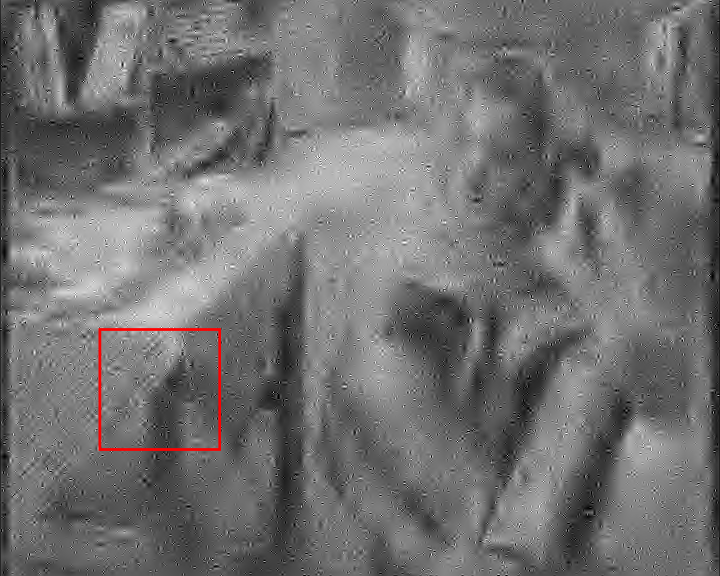}
	\end{subfigure}
    \put(-70,-23){\rotatebox{90}{MR = 0.05}}
	\vspace{-7.1mm}
	\begin{subfigure}[t]{0.115\textwidth}
		\centering
		\includegraphics[width=\columnwidth]{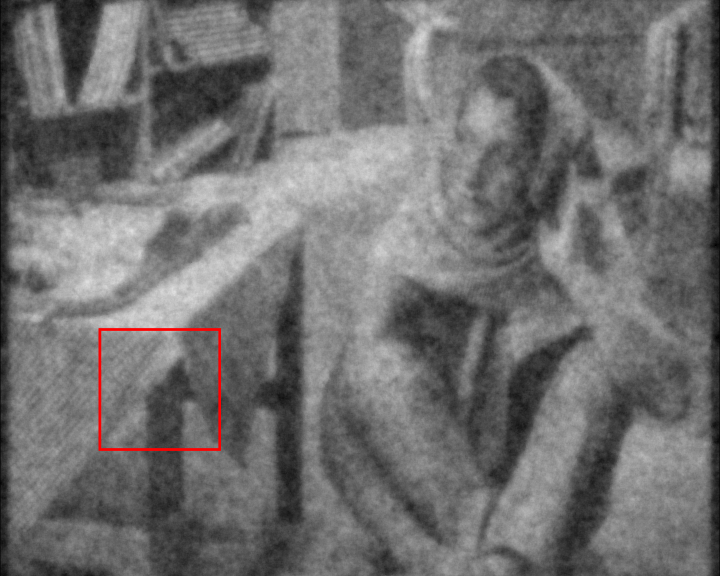}
	\end{subfigure}
	\begin{subfigure}[t]{0.115\textwidth}
		\centering
		\includegraphics[width=\columnwidth]{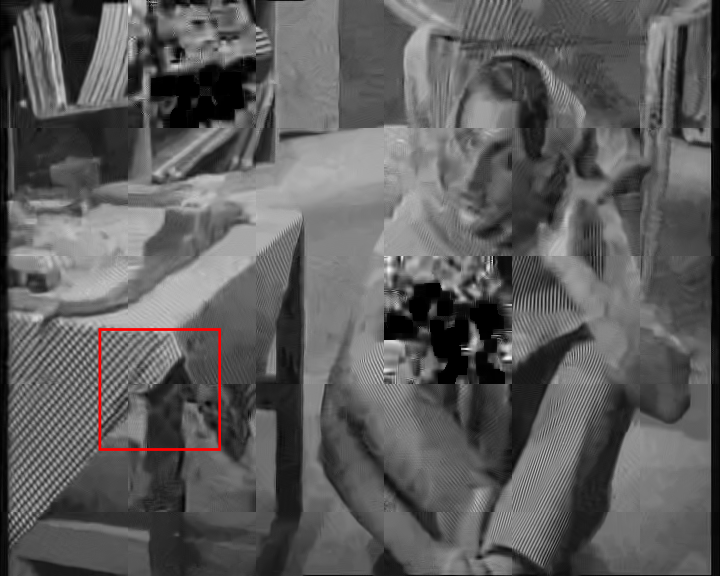}
	\end{subfigure}
	\begin{subfigure}[t]{0.115\textwidth}
		\centering
		\includegraphics[width=\columnwidth]{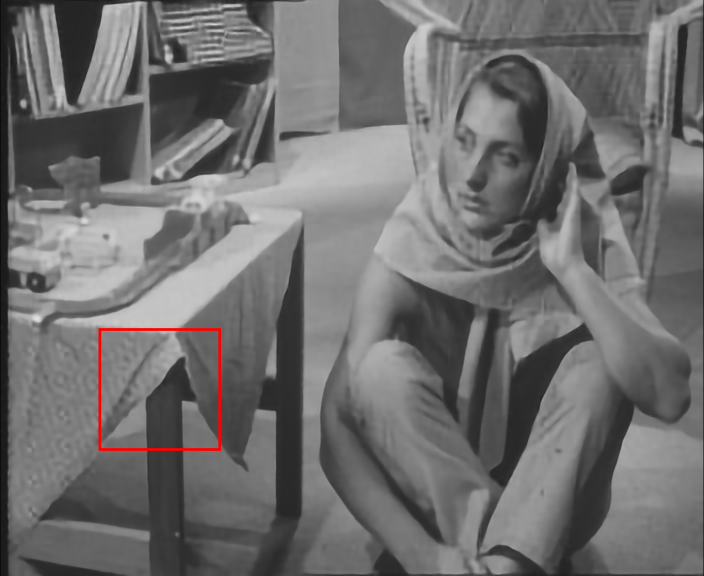}
	\end{subfigure}
	\begin{subfigure}[t]{0.115\textwidth}
		\centering
		\includegraphics[width=\columnwidth]{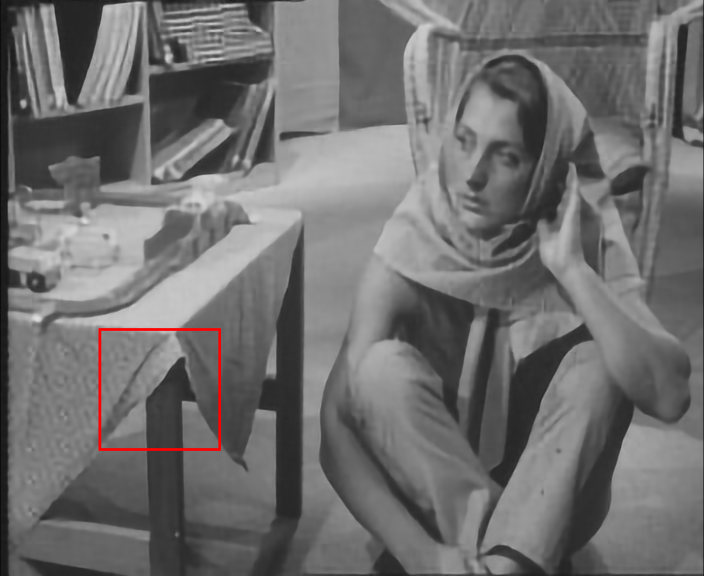}
	\end{subfigure}
	\begin{subfigure}[t]{0.115\textwidth}
		\centering
		\includegraphics[width=\columnwidth]{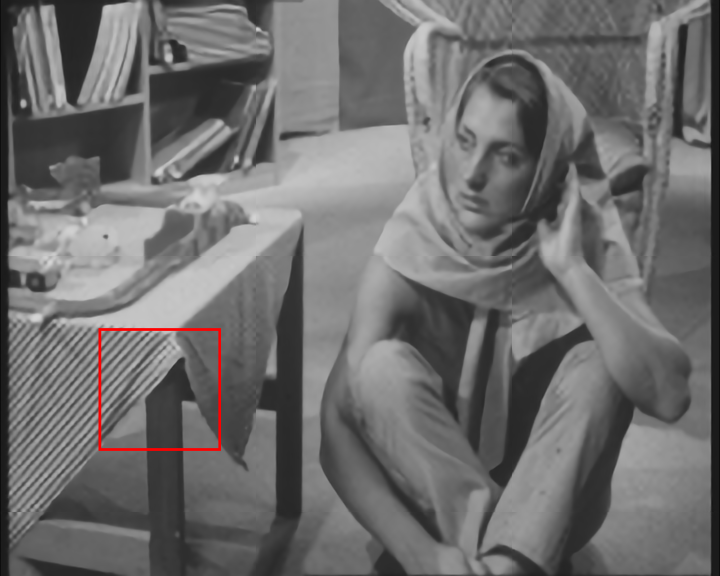}
	\end{subfigure}
	\begin{subfigure}[t]{0.115\textwidth}
		\centering
		\includegraphics[width=\columnwidth]{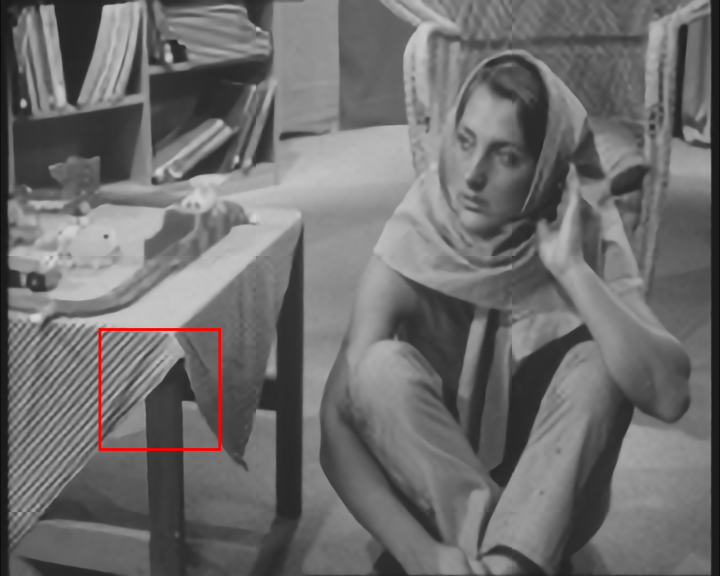}
	\end{subfigure}
	\begin{subfigure}[t]{0.115\textwidth}
		\centering
		\includegraphics[width=\columnwidth]{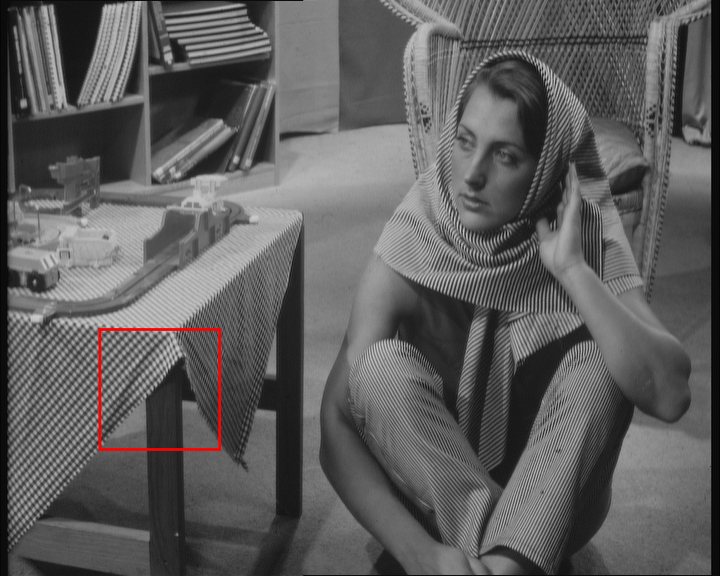}
	\end{subfigure}

	\begin{subfigure}[t]{0.115\textwidth}
		\centering
		\includegraphics[width=\columnwidth]{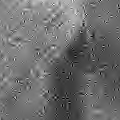}
		\caption{20.25dB}
	\end{subfigure}
	\vspace{1mm}	
	\begin{subfigure}[t]{0.115\textwidth}
		\centering
		\includegraphics[width=\columnwidth]{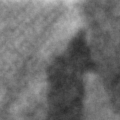}
		\caption{19.66dB}
	\end{subfigure}
	\begin{subfigure}[t]{0.115\textwidth}
		\centering
		\includegraphics[width=\columnwidth]{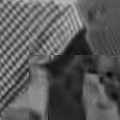}
		\caption{23.59dB}
	\end{subfigure}
	\begin{subfigure}[t]{0.115\textwidth}
		\centering
		\includegraphics[width=\columnwidth]{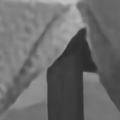}
		\caption{25.24dB}
	\end{subfigure}
	\begin{subfigure}[t]{0.115\textwidth}
		\centering
		\includegraphics[width=\columnwidth]{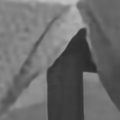}
		\caption{25.25dB}
	\end{subfigure}
	\begin{subfigure}[t]{0.115\textwidth}
		\centering
		\includegraphics[width=\columnwidth]{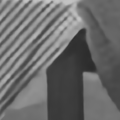}
		\caption{25.66dB}
	\end{subfigure}
	\begin{subfigure}[t]{0.115\textwidth}
		\centering
		\includegraphics[width=\columnwidth]{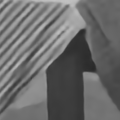}
		\caption{25.65dB}
	\end{subfigure}
	\begin{subfigure}[t]{0.115\textwidth}
		\centering
		\includegraphics[width=\columnwidth]{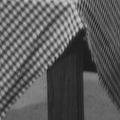}
		\caption{}
	\end{subfigure}
	\caption{Visual comparison with the state-of-the-art on grayscale images with varying measurement rates.}
	\label{figure:sotagray}
\end{figure*}

%% file: results/figure_compare_soa_DL_gray.tex
\begin{figure*}[htbp]	
	\centering
	\captionsetup[subfigure]{labelformat=empty}
	\begin{subfigure}[t]{0.115\textwidth}
		\caption{ReconNet}
		\centering
		\includegraphics[width=\columnwidth]{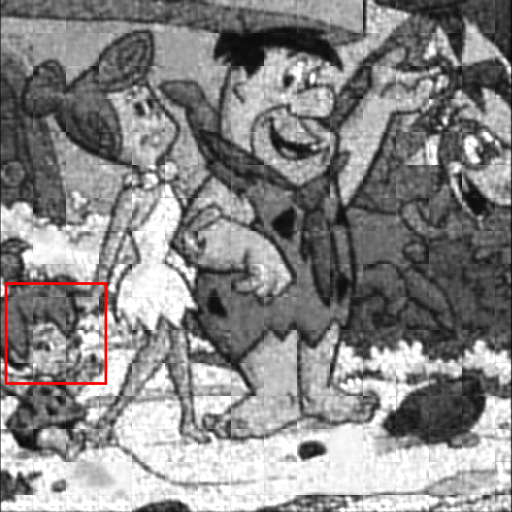}
	\end{subfigure}
	\put(-70,-98){\rotatebox{90}{MR = 0.1}}
	\vspace{-7mm}
	\begin{subfigure}[t]{0.115\textwidth}
		\caption{ISTA-Net+}
		\centering
		\includegraphics[width=\columnwidth]{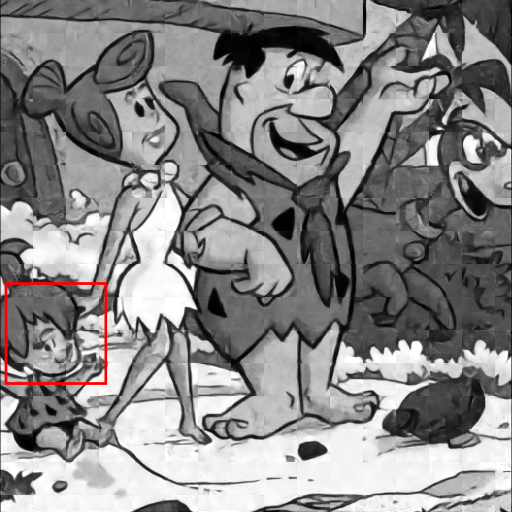}
	\end{subfigure}
	\begin{subfigure}[t]{0.115\textwidth}
		\caption{CSNet+}
		\centering
		\includegraphics[width=\columnwidth]{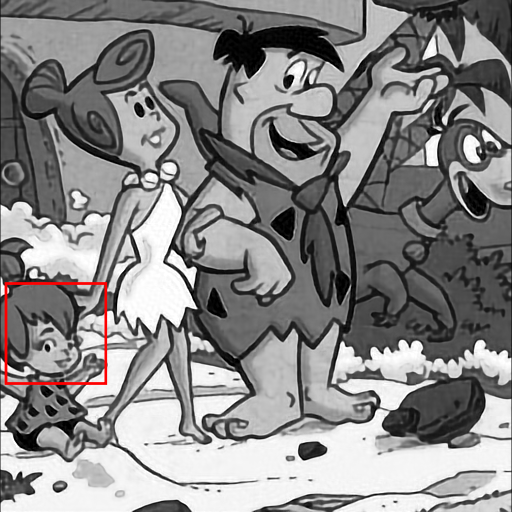}
	\end{subfigure}
	\begin{subfigure}[t]{0.115\textwidth}
		\caption{SCS-Net}
		\centering
		\includegraphics[width=\columnwidth]{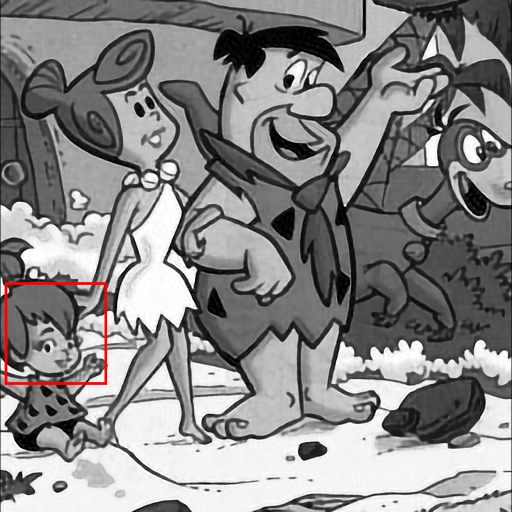}
	\end{subfigure}
	\begin{subfigure}[t]{0.115\textwidth}
		\caption{MAC}
		\centering
		\includegraphics[width=\columnwidth]{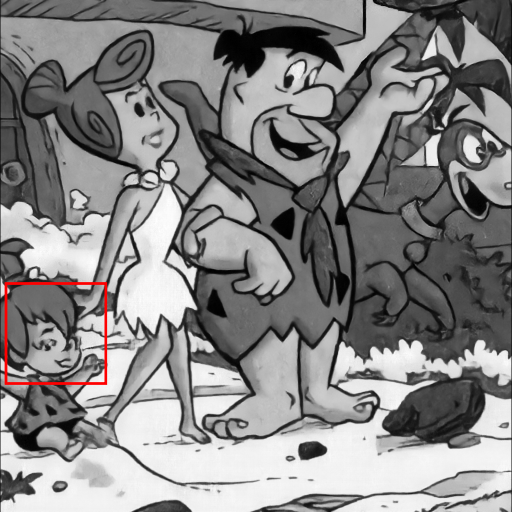}
	\end{subfigure}
	\begin{subfigure}[t]{0.115\textwidth}
		\caption{GTSNET-1}
		\centering
		\includegraphics[width=\columnwidth]{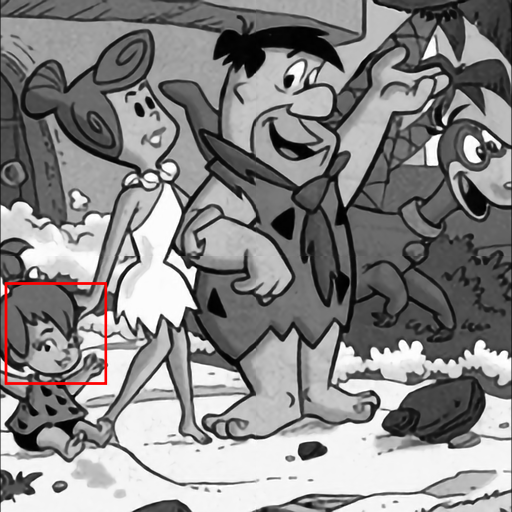}
	\end{subfigure}
	\begin{subfigure}[t]{0.115\textwidth}
		\caption{GTSNET-3}
		\centering
		\includegraphics[width=\columnwidth]{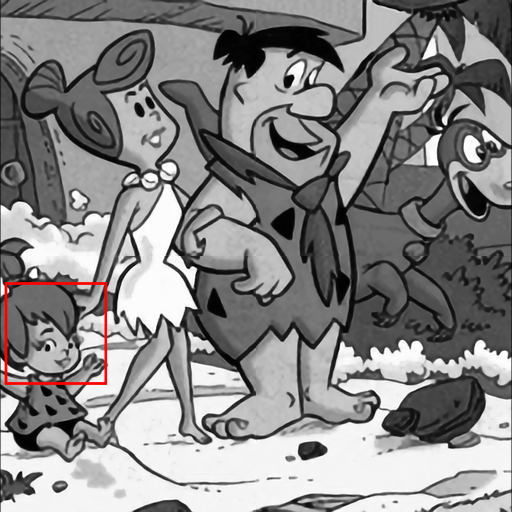}
	\end{subfigure}
	\begin{subfigure}[t]{0.115\textwidth}
		\caption{GT}
		\centering
		\includegraphics[width=\columnwidth]{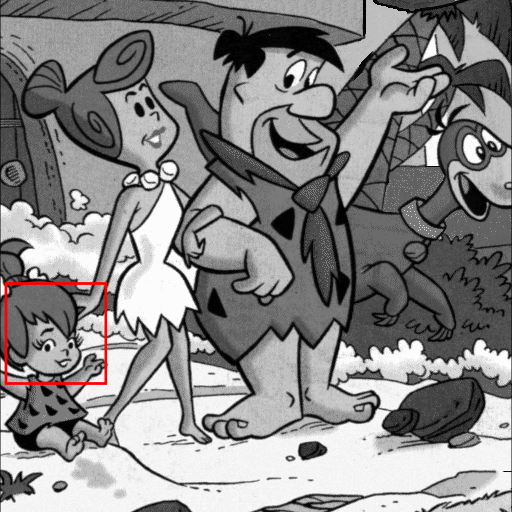}
	\end{subfigure}

	\begin{subfigure}[t]{0.115\textwidth}
		\centering
		\includegraphics[width=\columnwidth]{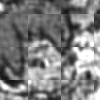}
		\caption{20.96dB}
	\end{subfigure}
	\vspace{1mm}	
	\begin{subfigure}[t]{0.115\textwidth}
		\centering
		\includegraphics[width=\columnwidth]{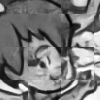}
		\caption{23.39dB}
	\end{subfigure}
	\begin{subfigure}[t]{0.115\textwidth}
		\centering
		\includegraphics[width=\columnwidth]{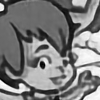}
		\caption{24.04dB}
	\end{subfigure}
	\begin{subfigure}[t]{0.115\textwidth}
		\centering
		\includegraphics[width=\columnwidth]{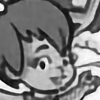}
		\caption{24.62dB}
	\end{subfigure}
	\begin{subfigure}[t]{0.115\textwidth}
		\centering
		\includegraphics[width=\columnwidth]{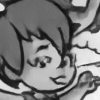}
		\caption{24.81dB}
	\end{subfigure}
	\begin{subfigure}[t]{0.115\textwidth}
		\centering
		\includegraphics[width=\columnwidth]{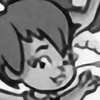}
		\caption{26.00dB}
	\end{subfigure}
	\begin{subfigure}[t]{0.115\textwidth}
		\centering
		\includegraphics[width=\columnwidth]{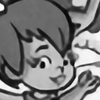}
		\caption{26.00dB}
	\end{subfigure}
	\begin{subfigure}[t]{0.115\textwidth}
		\centering
		\includegraphics[width=\columnwidth]{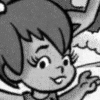}
		\caption{}
	\end{subfigure}
	
	\begin{subfigure}[t]{0.115\textwidth}
		\centering
		\includegraphics[width=\columnwidth]{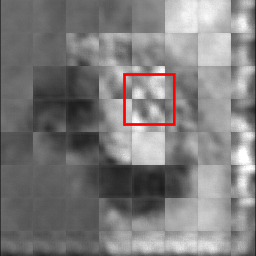}
	\end{subfigure}
    \put(-70,-23){\rotatebox{90}{MR = 0.01}}
	\vspace{-7mm}
	\begin{subfigure}[t]{0.115\textwidth}
		\centering
		\includegraphics[width=\columnwidth]{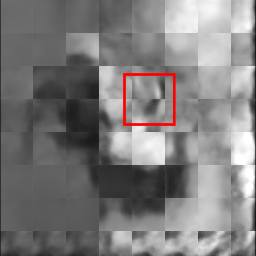}
	\end{subfigure}
	\begin{subfigure}[t]{0.115\textwidth}
		\centering
		\includegraphics[width=\columnwidth]{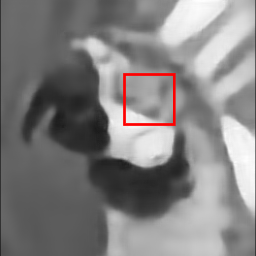}
	\end{subfigure}
	\begin{subfigure}[t]{0.115\textwidth}
		\centering
		\includegraphics[width=\columnwidth]{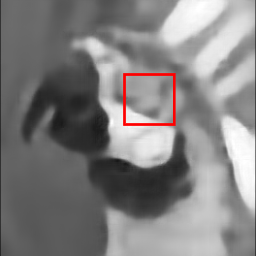}
	\end{subfigure}
	\begin{subfigure}[t]{0.115\textwidth}
		\centering
		\includegraphics[width=\columnwidth]{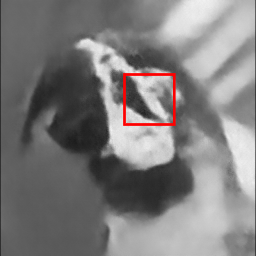}
	\end{subfigure}
	\begin{subfigure}[t]{0.115\textwidth}
		\centering
		\includegraphics[width=\columnwidth]{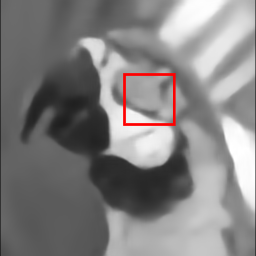}
	\end{subfigure}
	\begin{subfigure}[t]{0.115\textwidth}
		\centering
		\includegraphics[width=\columnwidth]{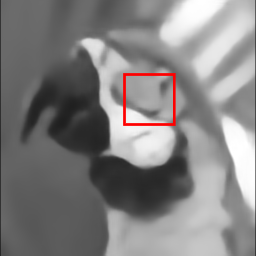}
	\end{subfigure}
	\begin{subfigure}[t]{0.115\textwidth}
		\centering
		\includegraphics[width=\columnwidth]{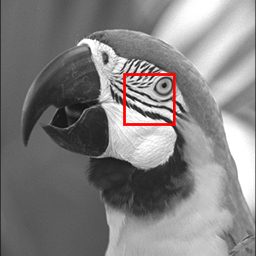}
	\end{subfigure}

	\begin{subfigure}[t]{0.115\textwidth}
		\centering
		\includegraphics[width=\columnwidth]{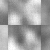}
		\caption{17.84dB}
	\end{subfigure}
	\vspace{1mm}	
	\begin{subfigure}[t]{0.115\textwidth}
		\centering
		\includegraphics[width=\columnwidth]{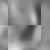}
		\caption{18.06dB}
	\end{subfigure}
	\begin{subfigure}[t]{0.115\textwidth}
		\centering
		\includegraphics[width=\columnwidth]{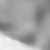}
		\caption{22.23dB}
	\end{subfigure}
	\begin{subfigure}[t]{0.115\textwidth}
		\centering
		\includegraphics[width=\columnwidth]{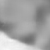}
		\caption{22.30dB}
	\end{subfigure}
	\begin{subfigure}[t]{0.115\textwidth}
		\centering
		\includegraphics[width=\columnwidth]{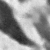}
		\caption{19.86dB}
	\end{subfigure}
	\begin{subfigure}[t]{0.115\textwidth}
		\centering
		\includegraphics[width=\columnwidth]{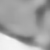}
		\caption{22.71dB}
	\end{subfigure}
	\begin{subfigure}[t]{0.115\textwidth}
		\centering
		\includegraphics[width=\columnwidth]{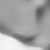}
		\caption{22.67dB}
	\end{subfigure}
	\begin{subfigure}[t]{0.115\textwidth}
		\centering
		\includegraphics[width=\columnwidth]{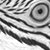}
		\caption{}
	\end{subfigure}
	
	\begin{subfigure}[t]{0.115\textwidth}
		\centering
		\includegraphics[width=\columnwidth]{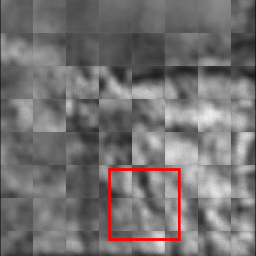}
	\end{subfigure}
    \put(-70,-23){\rotatebox{90}{MR = 0.01}}
	\vspace{-7mm}
	\begin{subfigure}[t]{0.115\textwidth}
		\centering
		\includegraphics[width=\columnwidth]{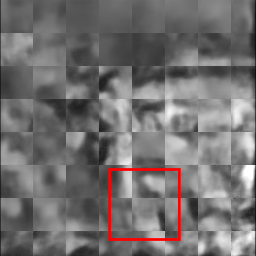}
	\end{subfigure}
	\begin{subfigure}[t]{0.115\textwidth}
		\centering
		\includegraphics[width=\columnwidth]{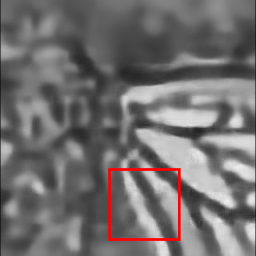}
	\end{subfigure}
	\begin{subfigure}[t]{0.115\textwidth}
		\centering
		\includegraphics[width=\columnwidth]{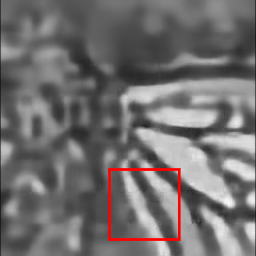}
	\end{subfigure}
	\begin{subfigure}[t]{0.115\textwidth}
		\centering
		\includegraphics[width=\columnwidth]{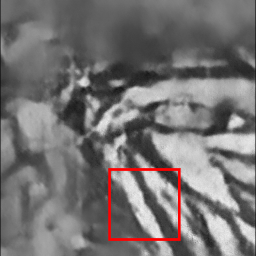}
	\end{subfigure}
	\begin{subfigure}[t]{0.115\textwidth}
		\centering
		\includegraphics[width=\columnwidth]{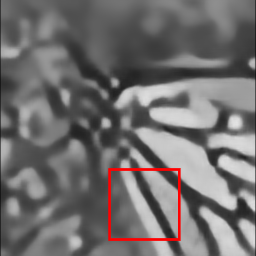}
	\end{subfigure}
	\begin{subfigure}[t]{0.115\textwidth}
		\centering
		\includegraphics[width=\columnwidth]{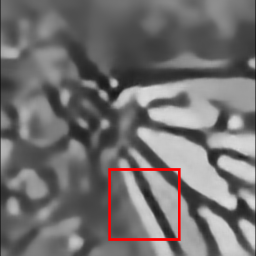}
	\end{subfigure}
	\begin{subfigure}[t]{0.115\textwidth}
		\centering
		\includegraphics[width=\columnwidth]{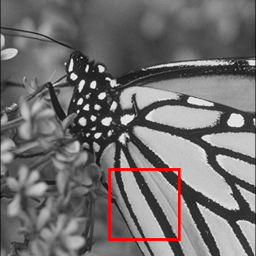}
	\end{subfigure}

	\begin{subfigure}[t]{0.115\textwidth}
		\centering
		\includegraphics[width=\columnwidth]{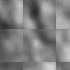}
		\caption{15.09dB}
	\end{subfigure}
	\vspace{1mm}	
	\begin{subfigure}[t]{0.115\textwidth}
		\centering
		\includegraphics[width=\columnwidth]{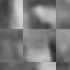}
		\caption{15.01dB}
	\end{subfigure}
	\begin{subfigure}[t]{0.115\textwidth}
		\centering
		\includegraphics[width=\columnwidth]{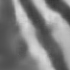}
		\caption{18.07dB}
	\end{subfigure}
	\begin{subfigure}[t]{0.115\textwidth}
		\centering
		\includegraphics[width=\columnwidth]{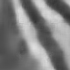}
		\caption{18.05dB}
	\end{subfigure}
	\begin{subfigure}[t]{0.115\textwidth}
		\centering
		\includegraphics[width=\columnwidth]{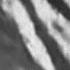}
		\caption{15.95dB}
	\end{subfigure}
	\begin{subfigure}[t]{0.115\textwidth}
		\centering
		\includegraphics[width=\columnwidth]{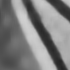}
		\caption{18.97dB}
	\end{subfigure}
	\begin{subfigure}[t]{0.115\textwidth}
		\centering
		\includegraphics[width=\columnwidth]{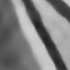}
		\caption{18.98dB}
	\end{subfigure}
	\begin{subfigure}[t]{0.115\textwidth}
		\centering
		\includegraphics[width=\columnwidth]{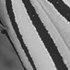}
		\caption{}
	\end{subfigure}	
	\caption{The recovered images of the competing and proposed methods with the GT image on the right.}
	\label{figure:deepmethods}
\end{figure*}

%% file: results/Plot_PSNRs_Set5_gray.tex
%
%
\definecolor{mycolor1}{rgb}{0.00000,0.44700,0.74100}%
\definecolor{mycolor2}{rgb}{0.85000,0.32500,0.09800}%
\definecolor{mycolor3}{rgb}{0.92900,0.69400,0.12500}%
\begin{tikzpicture}

\begin{axis}[%
width=0.951\columnwidth,
height=0.893\columnwidth,
at={(0\columnwidth,0\columnwidth)},
scale only axis,
xmin=1,
xmax=5,
xlabel style={font=\color{white!15!black},yshift=3pt},
xtick={1,2,3,4,5},
xticklabels={baby,bird,butterfly,head,woman},
x tick label style={rotate=45},
ymin=28,
ymax=36,
ylabel style={font=\color{white!15!black},yshift=-3pt},
ylabel={PSNR},
axis background/.style={fill=white},
title style={font=\bfseries},
title={Grayscale},
legend style={legend cell align=left, align=left,
at={(0,0)},anchor=south west,
nodes={scale=0.7, transform shape},
draw=white!15!black}
]
\addplot [color=mycolor1, line width=1.4pt]
  table[row sep=crcr]{%
1	35.1647529602051\\
2	35.5974578857422\\
3	28.9138431549072\\
4	33.7916526794434\\
5	31.7020397186279\\
};
\addlegendentry{$T=1$}

\addplot [color=mycolor2, line width=1.4pt]
  table[row sep=crcr]{%
1	35.1715698242188\\
2	35.9786605834961\\
3	28.8895225524902\\
4	33.7372817993164\\
5	31.7568511962891\\
};
\addlegendentry{$T=3$}

\addplot [color=mycolor3, line width=1.4pt]
  table[row sep=crcr]{%
1	35.1631469726563\\
2	35.8560562133789\\
3	29.0089054107666\\
4	33.7205696105957\\
5	31.7590522766113\\
};
\addlegendentry{$T=5$}

\end{axis}

\begin{axis}[%
width=1.227\columnwidth,
height=1.095\columnwidth,
at={(-0.16\columnwidth,-0.12\columnwidth)},
scale only axis,
xmin=0,
xmax=1,
ymin=0,
ymax=1,
axis line style={draw=none},
ticks=none,
axis x line*=bottom,
axis y line*=left,
legend style={legend cell align=left, align=left, draw=white!15!black}
]
\end{axis}
\end{tikzpicture}%

%% file: results/Plot_PSNRs_Set5_rgb.tex
%
%
\definecolor{mycolor1}{rgb}{0.00000,0.44700,0.74100}%
\definecolor{mycolor2}{rgb}{0.85000,0.32500,0.09800}%
\definecolor{mycolor3}{rgb}{0.92900,0.69400,0.12500}%
\begin{tikzpicture}

\begin{axis}[%
width=0.951\columnwidth,
height=0.893\columnwidth,
at={(0\columnwidth,0\columnwidth)},
scale only axis,
xmin=1,
xmax=5,
xlabel style={font=\color{white!15!black},yshift=3pt},
xtick={1,2,3,4,5},
xticklabels={baby,bird,butterfly,head,woman},
x tick label style={rotate=45},
ymin=27,
ymax=36,
ylabel style={font=\color{white!15!black},yshift=-3pt},
ylabel={PSNR},
axis background/.style={fill=white},
title style={font=\bfseries},
title={RGB},
legend style={legend cell align=left, align=left,
at={(0,0)},anchor=south west,
nodes={scale=0.7, transform shape},
draw=white!15!black}
]
\addplot [color=mycolor1, line width=1.4pt]
  table[row sep=crcr]{%
1	33.7636642456055\\
2	33.695743560791\\
3	27.5297527313232\\
4	30.566291809082\\
5	30.2515773773193\\
};
\addlegendentry{$T=1$}

\addplot [color=mycolor2, line width=1.4pt]
  table[row sep=crcr]{%
1	35.713924407959\\
2	35.014274597168\\
3	29.9021301269531\\
4	31.2400512695313\\
5	32.9927024841309\\
};
\addlegendentry{$T=3$}

\addplot [color=mycolor3, line width=1.4pt]
  table[row sep=crcr]{%
1	35.4465217590332\\
2	34.3462905883789\\
3	29.0651092529297\\
4	31.0428771972656\\
5	32.3320617675781\\
};
\addlegendentry{$T=5$}

\end{axis}

\begin{axis}[%
width=1.227\columnwidth,
height=1.095\columnwidth,
at={(-0.16\columnwidth,-0.12\columnwidth)},
scale only axis,
xmin=0,
xmax=1,
ymin=0,
ymax=1,
axis line style={draw=none},
ticks=none,
axis x line*=bottom,
axis y line*=left,
legend style={legend cell align=left, align=left, draw=white!15!black}
]
\end{axis}
\end{tikzpicture}%

%% file: results/figure_freq_test_rgb.tex
\begin{figure}[htbp]	
	\centering
	\captionsetup[subfigure]{labelformat=empty}
	\begin{subfigure}[t]{0.115\textwidth}
		\caption{GTSNET-1}
		\centering
		\includegraphics[width=\columnwidth]{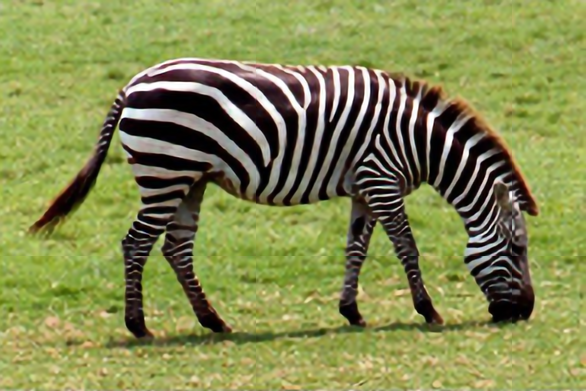}
	\end{subfigure}
	\vspace{1mm}
	\begin{subfigure}[t]{0.115\textwidth}
		\caption{GTSNET-3}
		\centering
		\includegraphics[width=\columnwidth]{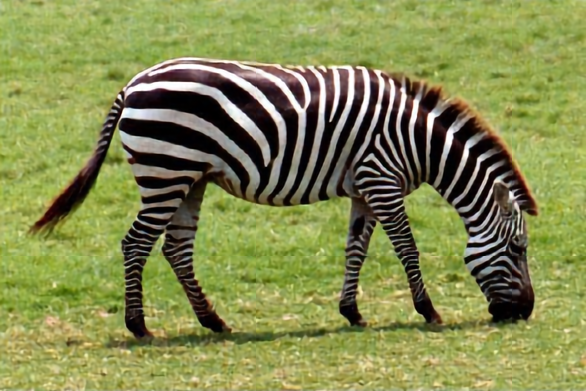}
	\end{subfigure}
	\begin{subfigure}[t]{0.115\textwidth}
		\caption{GTSNET-5}
		\centering
		\includegraphics[width=\columnwidth]{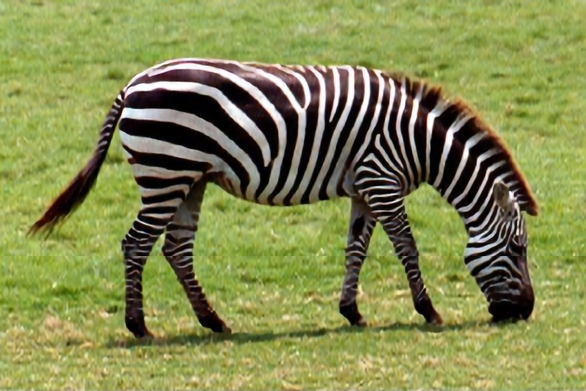}
	\end{subfigure}
	\begin{subfigure}[t]{0.115\textwidth}
		\caption{GT}
		\centering
		\includegraphics[width=\columnwidth]{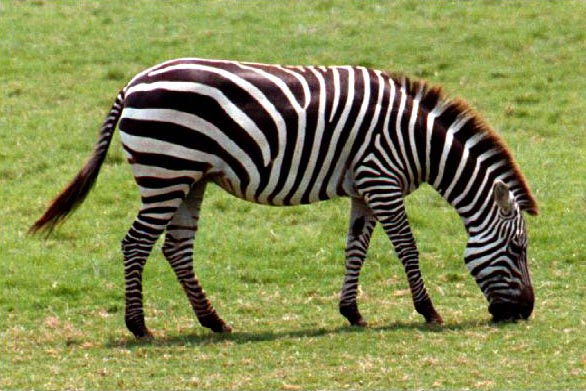}
	\end{subfigure}

    \begin{subfigure}[t]{0.115\textwidth}
		\centering
		\includegraphics[width=\columnwidth]{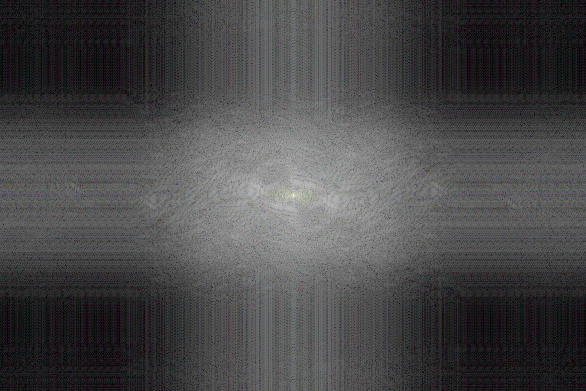}
	\end{subfigure}
	\vspace{1mm}
	\begin{subfigure}[t]{0.115\textwidth}
		\centering
		\includegraphics[width=\columnwidth]{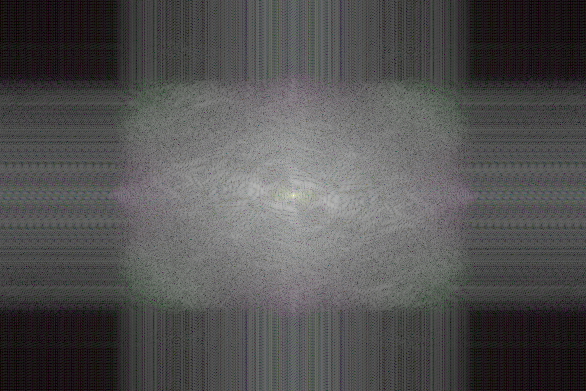}
	\end{subfigure}
	\begin{subfigure}[t]{0.115\textwidth}
		\centering
		\includegraphics[width=\columnwidth]{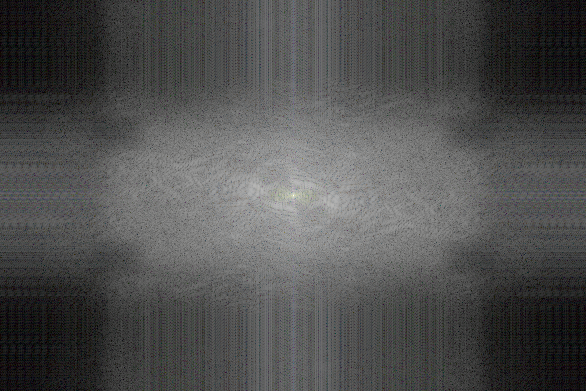}
	\end{subfigure}
	\begin{subfigure}[t]{0.115\textwidth}
		\centering
		\includegraphics[width=\columnwidth]{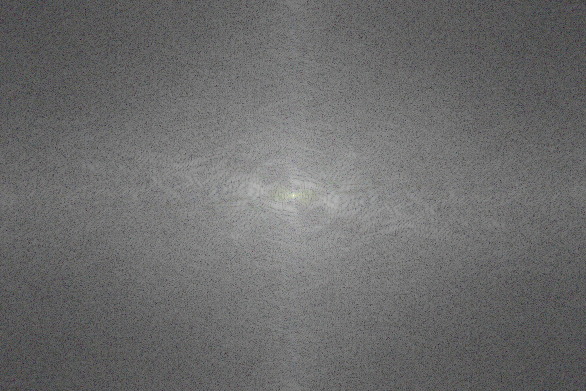}
	\end{subfigure}

	\begin{subfigure}[t]{0.115\textwidth}
		\centering
		\includegraphics[width=\columnwidth]{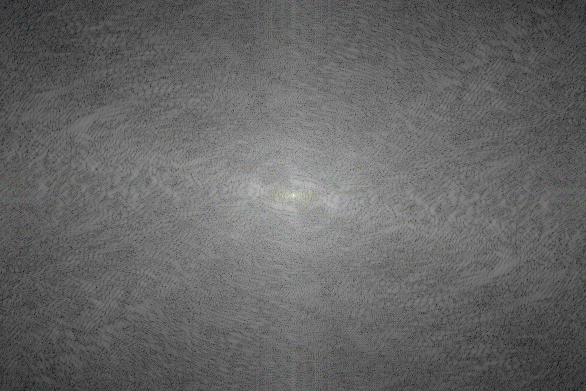}
		\caption{27.72dB}
	\end{subfigure}
	\begin{subfigure}[t]{0.115\textwidth}
		\centering
		\includegraphics[width=\columnwidth]{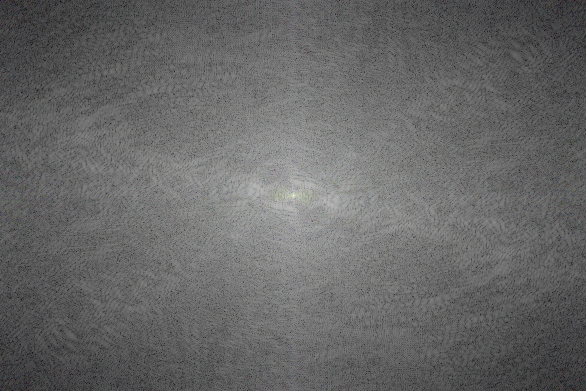}
		\caption{30.85dB}
	\end{subfigure}
	\begin{subfigure}[t]{0.115\textwidth}
		\centering
		\includegraphics[width=\columnwidth]{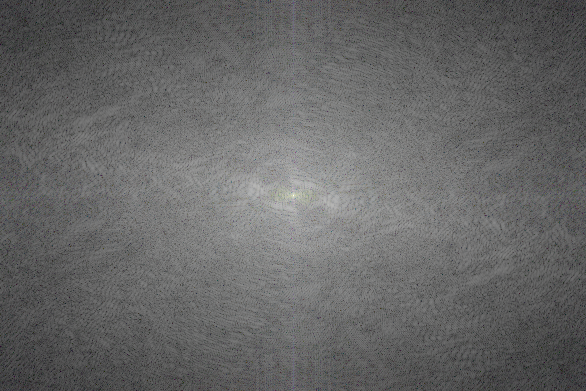}
		\caption{29.13dB}
	\end{subfigure}
	\begin{subfigure}[t]{0.115\textwidth}
		\centering
		\includegraphics[width=\columnwidth]{results/freqTests/rgb/zebra_gt_FFT.png}
		\caption{}
	\end{subfigure}
	\caption{Frequency analysis with varying number of tensor sums. Top: Final outputs $\mathcal{\widehat{S}}$ in spatial domain. Middle: Proxy signals $\mathcal{\widetilde{S}}$ in frequency domain. Bottom: Final outputs $\mathcal{\widehat{S}}$ in frequency domain. Measurement rate is set as 0.1.}
	\label{fig:freq_tests_rgb}
\end{figure}

%% file: results/figure_GTSNET3_Proxy_Branches.tex
\begin{figure}[htbp]	
	\centering
	\captionsetup[subfigure]{labelformat=empty}
	\begin{subfigure}[t]{0.09\textwidth}
		\caption{$\mathbf{B^{(1)}}\mathcal{Y}$}
		\centering
		\includegraphics[width=\columnwidth]{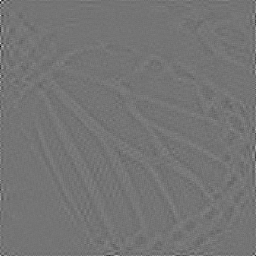}
	\end{subfigure}
	\vspace{1mm}
	\begin{subfigure}[t]{0.09\textwidth}
		\caption{$\mathbf{B^{(2)}}\mathcal{Y}$}
		\centering
		\includegraphics[width=\columnwidth]{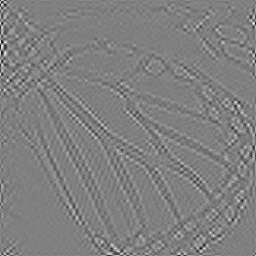}
	\end{subfigure}
	\begin{subfigure}[t]{0.09\textwidth}
		\caption{$\mathbf{B^{(3)}}\mathcal{Y}$}
		\centering
		\includegraphics[width=\columnwidth]{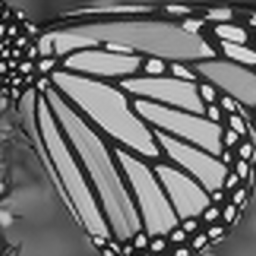}
	\end{subfigure}
	\begin{subfigure}[t]{0.09\textwidth}
	    \vspace{-0.2mm}
		\caption{$\mathcal{\widetilde{S}}$}
		\centering
		\includegraphics[width=\columnwidth]{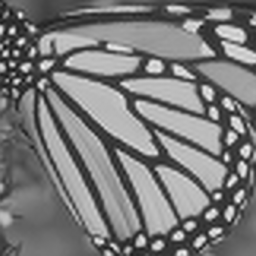}
	\end{subfigure}
	\begin{subfigure}[t]{0.09\textwidth}
	    \vspace{0.45mm}
		\caption{GT}
		\centering
		\includegraphics[width=\columnwidth]{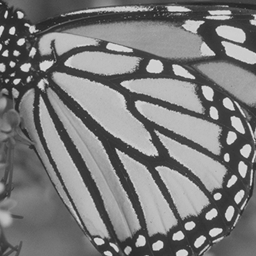}
	\end{subfigure}

    \begin{subfigure}[t]{0.09\textwidth}
		\centering
		\includegraphics[width=\columnwidth]{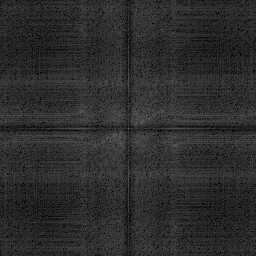}
	\end{subfigure}
	\begin{subfigure}[t]{0.09\textwidth}
		\centering
		\includegraphics[width=\columnwidth]{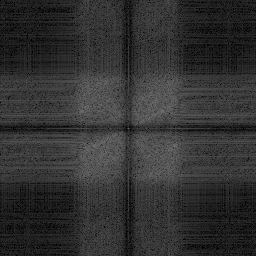}
	\end{subfigure}
	\begin{subfigure}[t]{0.09\textwidth}
		\centering
		\includegraphics[width=\columnwidth]{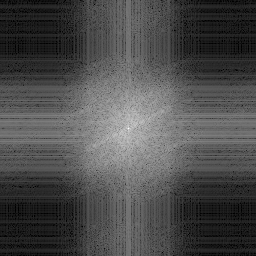}
	\end{subfigure}
	\begin{subfigure}[t]{0.09\textwidth}
		\centering
		\includegraphics[width=\columnwidth]{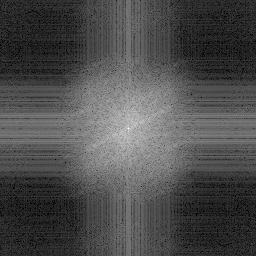}
	\end{subfigure}
	\begin{subfigure}[t]{0.09\textwidth}
		\centering
		\includegraphics[width=\columnwidth]{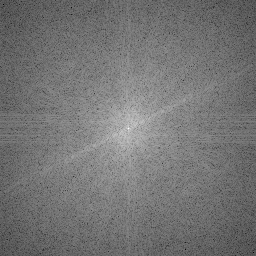}
	\end{subfigure}

	\caption{Tentative reconstruction result of each individual adjoint operation in GTSNET-3, as well as the proxy signal $\mathcal{\widetilde{S}}$ as the summation of each branch.}
	\label{fig:Proxy_Branches}
\end{figure}

%% file: results/figure_NoDCT_vs_DCT_rgb.tex
\begin{figure}[htbp]	
	\centering
	\captionsetup[subfigure]{labelformat=empty}
	\begin{subfigure}[t]{0.155\textwidth}
		\caption{Unstructured}
		\centering
		\includegraphics[width=\columnwidth]{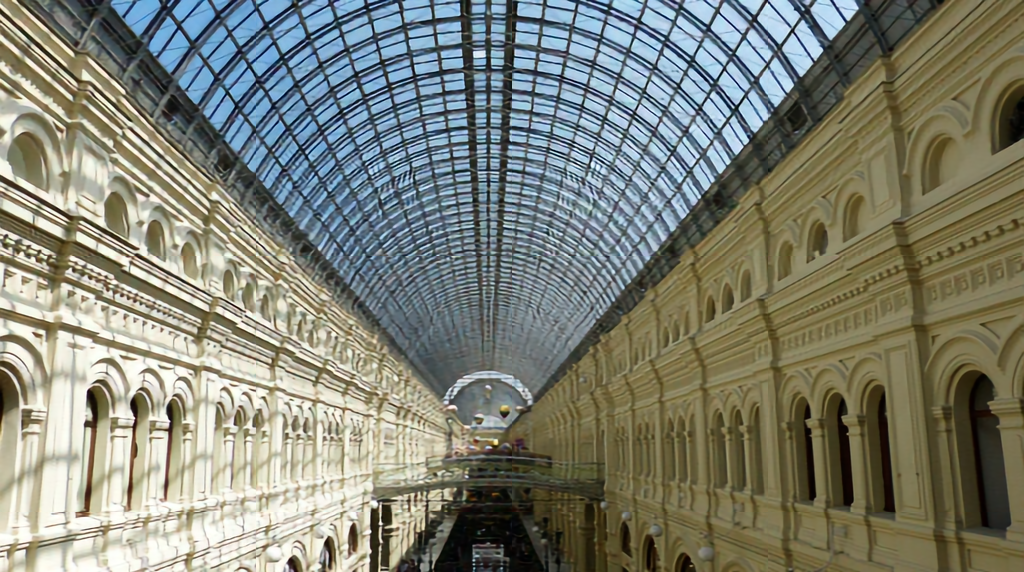}
	\end{subfigure}
	\vspace{1mm}
	\begin{subfigure}[t]{0.155\textwidth}
		\caption{DCT}
		\centering
		\includegraphics[width=\columnwidth]{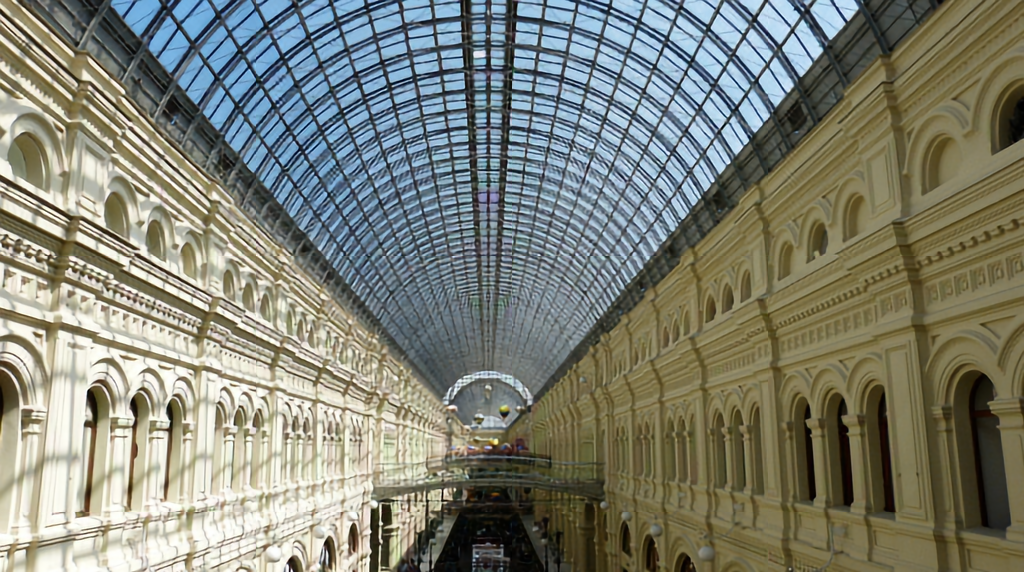}
	\end{subfigure}
	\begin{subfigure}[t]{0.155\textwidth}
		\caption{GT}
		\centering
		\includegraphics[width=\columnwidth]{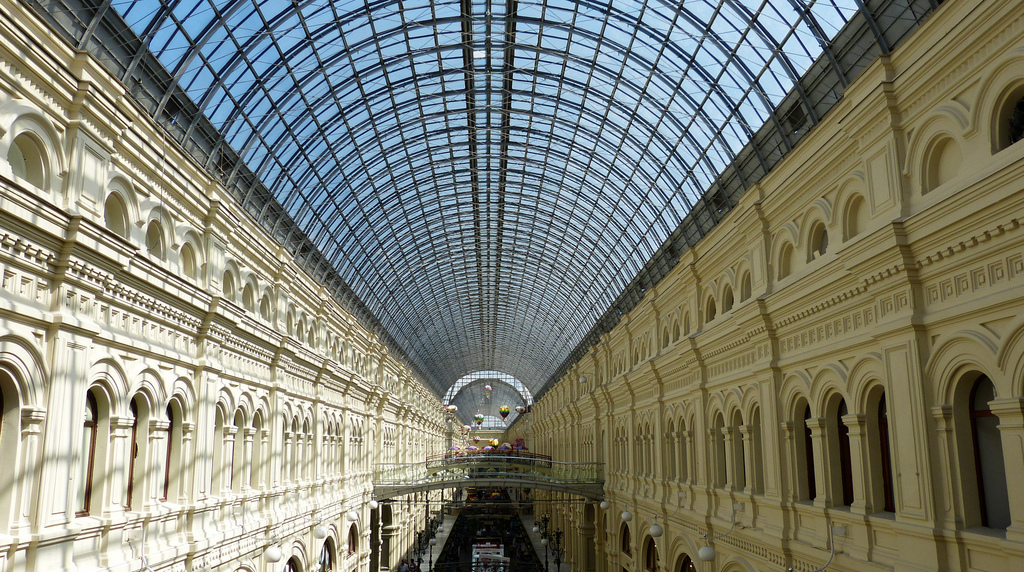}
	\end{subfigure}

    \begin{subfigure}[t]{0.155\textwidth}
		\centering
		\includegraphics[width=\columnwidth]{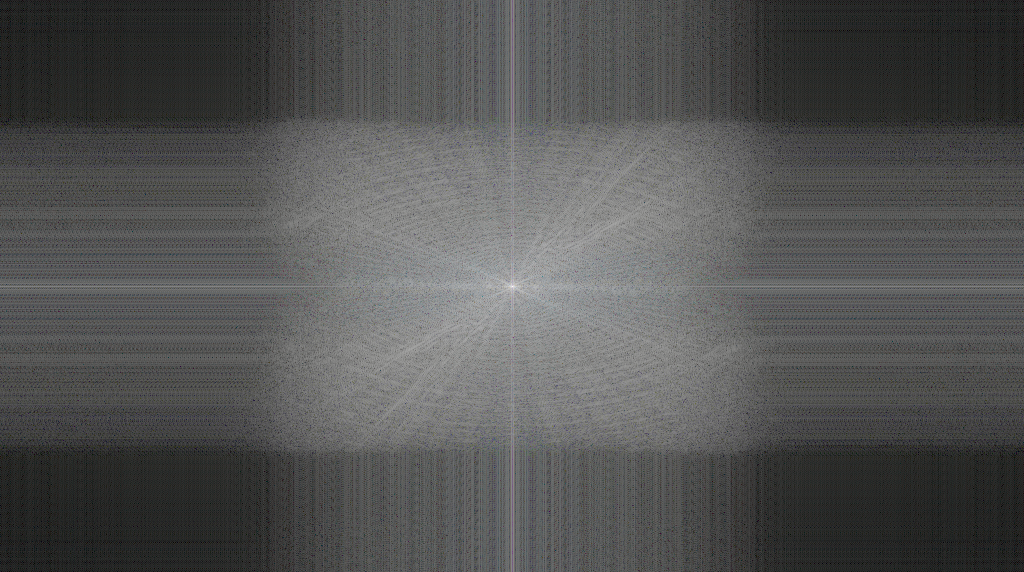}
	\end{subfigure}
	\vspace{1mm}
	\begin{subfigure}[t]{0.155\textwidth}
		\centering
		\includegraphics[width=\columnwidth]{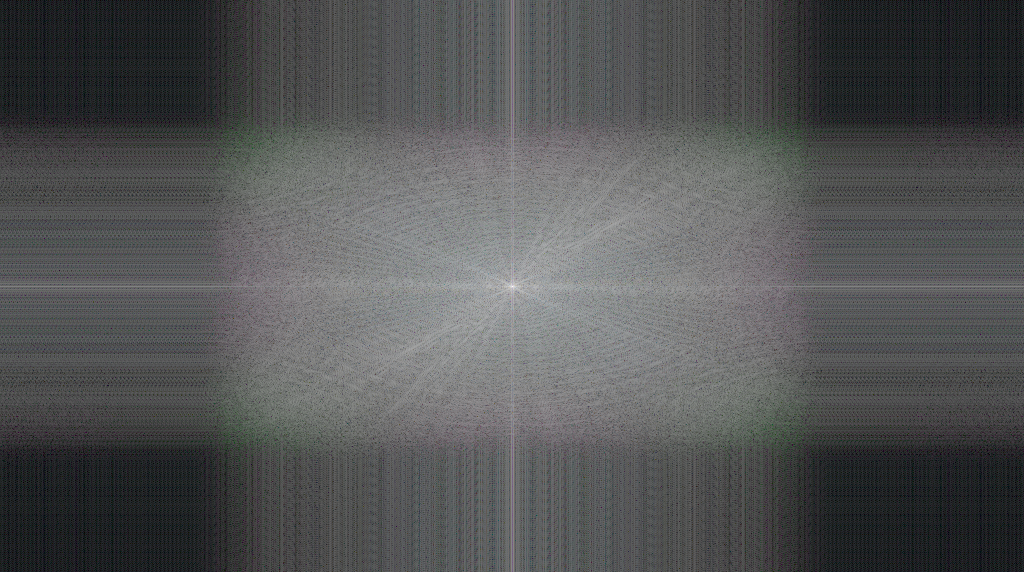}
	\end{subfigure}
	\begin{subfigure}[t]{0.155\textwidth}
		\centering
		\includegraphics[width=\columnwidth]{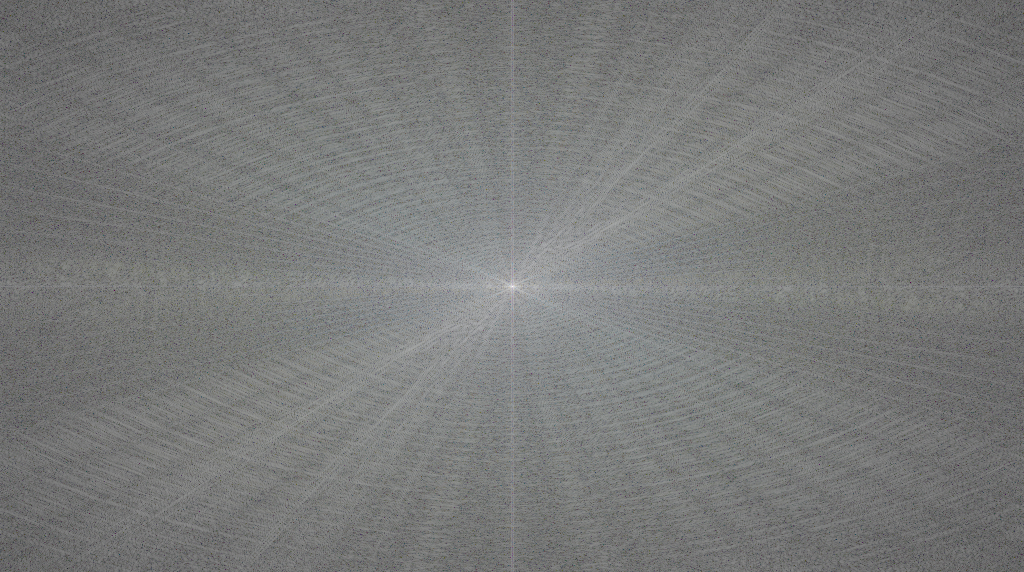}
	\end{subfigure}

	 \begin{subfigure}[t]{0.155\textwidth}
		\centering
		\includegraphics[width=\columnwidth]{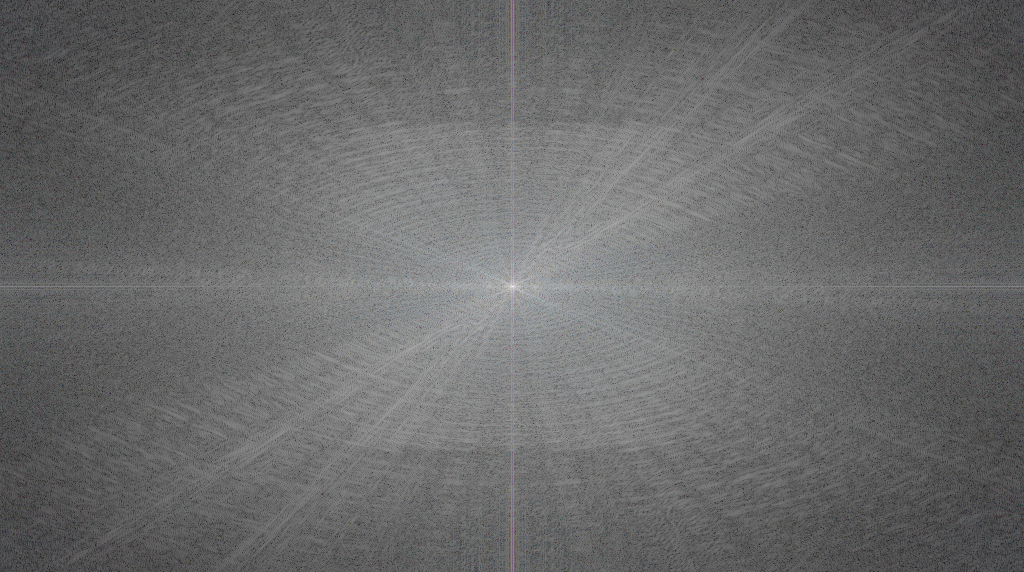}
		\caption{22.28dB}
	\end{subfigure}
	\begin{subfigure}[t]{0.155\textwidth}
		\centering
		\includegraphics[width=\columnwidth]{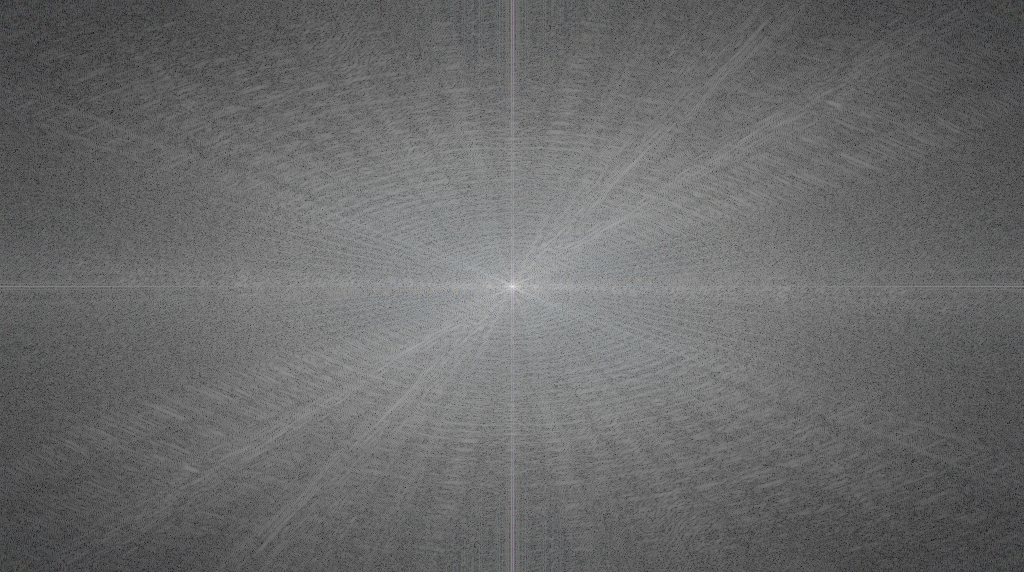}
		\caption{22.71dB}
	\end{subfigure}
	\begin{subfigure}[t]{0.155\textwidth}
		\centering
		\includegraphics[width=\columnwidth]{results/freqTests/NoDCT_rgb/img_008_gt_FFT.png}
		\caption{}
	\end{subfigure}
	\caption{Visual comparison between the unstructured and structured (DCT) Tensor sums. Top: Final outputs $\mathcal{\widehat{S}}$ in the spatial domain. Middle: Proxy signals $\mathcal{\widetilde{S}}$ in the frequency domain. Bottom: Final outputs $\mathcal{\widehat{S}}$ in the frequency domain. The results are shown for GTSNET-3 with MR=0.1.}
	\label{fig:NoDCT_vs_DCT}
\end{figure}